\documentclass[10pt,reqno]{amsart}
\usepackage{a4wide,graphicx}

\usepackage{amssymb, amsbsy, amsfonts}
\usepackage{amsmath, amsthm}

\usepackage{color}

\usepackage{hyperref}

\usepackage{soul}

\allowdisplaybreaks[3]

\newcommand{\SigmaP}{{\tt Sigma}}

\newcounter{linectr}
\newenvironment{alg}[5]
{\begin{Algorithm}#1\makebox[0.0cm][l]{}\small{#2}\\
		\vspace*{0.0cm}\noindent\hspace*{0.1cm}\makebox[1.3cm][l]{\sf
			Input:}\begin{minipage}[t]{12.8cm}{#3}\end{minipage}\\
		\vspace*{0.1cm}\noindent\hspace*{0.1cm}\makebox[1.3cm][l]{\sf
			Output:}\begin{minipage}[t]{12.8cm}{#4}\end{minipage}\\
		\vspace*{-0.4cm}
		\begin{list}{(\arabic{linectr})}{\usecounter{linectr} \labelwidth3ex\itemsep0ex\labelsep1ex\leftmargin5ex\parskip-0.2cm
				\listparindent0ex}}
		{\end{list}\end{Algorithm}\normalsize}

\newcommand{\xFy}[5]{{_{#1} F_{#2}} \left( \genfrac{}{}{0pt}{0}{#3}{#4} ; {#5} \right) }

\newcommand{\xFyk}[5]{{_{#1} F_{#2}} \left( \genfrac{}{}{0pt}{0}{#3}{#4} ; {#5} \right)_k }

\RequirePackage{url}

\newcommand{\QQ}{\mathbb{Q}}
\newcommand{\FF}{\mathbb{F}}
\newcommand{\KK}{\mathbb{K}}
\newcommand{\CC}{\mathbb{C}}
\newcommand{\NN}{\mathbb{N}}
\newcommand{\ZZ}{\mathbb{Z}}
\newcommand{\lcm}{{\rm lcm}}

\newcommand{\notion}[1]{{\em #1}}

\theoremstyle{plain}
\newtheorem{Theorem}{Theorem}

\theoremstyle{definition}

\newtheorem{rem}{Remark}
\newtheorem{exam}{Example}
\newtheorem{Algorithm}{Method}

\newenvironment{Example}{\begin{exam}}{\qed\end{exam}}
\newenvironment{Remark}{\begin{rem}}{
		\qed\end{rem}}

\definecolor{blaugrau}{rgb}{0.796887, 0.789075, 0.871107}
\newcounter{mmacnt}
\def\restartmma{\setcounter{mmacnt}{0}}
\restartmma \catcode`|=\active
\def|#1|{\mathrm{#1}}
\catcode`|=12
\newenvironment{mma}{
	\par
	\catcode`|=\active
	\parskip=4pt\parindent=0pt 
	\small
	\def\In##1\\{%
		\def\linebreak{\hfill\break\null\qquad}%
		\refstepcounter{mmacnt}
		\hangindent=2.5em\hangafter=0
		\leavevmode
		\llap{\tiny\sffamily In[\arabic{mmacnt}]:=\kern.5em}%
		\mathversion{bold}\footnotesize$\tt\bf\displaystyle##1$\normalsize
		\mathversion{normal}\par
	}%
	\def\Print##1\\{%
		\def\linebreak{\hfill\break}%
		\hangindent=2.5em\hangafter=0
		\leavevmode\scriptsize ##1\par}%
	\def\Out##1\\{%
		\vspace*{-0.2cm}\def\linebreak{$\hfill\break\null\hfill$}%
		\kern\abovedisplayskip\par
		\hangindent=2.5em\hangafter=0
		\leavevmode
		\llap{\tiny\sffamily Out[\arabic{mmacnt}]=\kern.5em}
		\footnotesize$\displaystyle\tt##1$\normalsize\hfill\null\par
		\kern\belowdisplayskip\vspace*{-0.3cm}
	}%
	\def\Warning##1##2\\{%
		\def\linebreak{\hfill\break}%
		\hangindent=2.5em\hangafter=0
		\leavevmode
		{\scriptsize##1 : ##2}\par}%
}{%
	\par\medskip
}

\newcommand{\LoadP}[1]{\fcolorbox{black}{blaugrau}{
		\begin{minipage}[t]{13cm}
			\footnotesize #1
\end{minipage}}}

\newcommand{\myIn}[1]{{\sffamily In[#1]}}
\newcommand{\myOut}[1]{{\sffamily Out[#1]}}

\def\MLabel#1{{\refstepcounter{mmacnt}\label{#1}}\addtocounter{mmacnt}{-1}}

\newcommand{\MText}[1]{\textbf{\ttfamily#1}}

\begin{document}

\author[P.~Paule]{Peter Paule}
\address[Peter~Paule]{
        Research Institute for Symbolic Computation (RISC)\\
        Johannes Kepler University Linz\\
        Altenbergerstr. 69\\
        A--4040 Linz, Austria\\
        and Center for Applied Mathematics,
        Tianjin University\\
        Tianjin 300072, P.R. China}
\email{Peter.Paule@risc.jku.at}

\author[C.~Schneider]{Carsten Schneider}
\address[Carsten~Schneider]{
        Research Institute for Symbolic Computation (RISC)\\
        Johannes Kepler University Linz\\
        Altenbergerstr. 69\\
        A--4040 Linz, Austria}
\email{Carsten.Schneider@risc.jku.at}
\thanks{This work was supported by the Austrian Science Fund (FWF) grant P33530.}

\begin{flushleft}
	RISC Report Series 24-01 
	\\[1cm]
\end{flushleft}

\title{Creative Telescoping for Hypergeometric Double Sums}

\begin{abstract}
We present efficient methods for calculating linear recurrences of hypergeometric double sums and, more generally, of multiple sums. In particular, we supplement this approach with the algorithmic theory of contiguous relations, which guarantees the applicability of our method for many input sums. In addition, we elaborate new techniques to optimize the underlying key task of our method to compute rational solutions of parameterized linear recurrences.
\end{abstract}

\dedicatory{Dedicated to the memory of 
our friend Marko Petkov{\v s}ek}

\maketitle

\section{Introduction}\label{Sec:Introduction}

We are interested in the following summation problem. {\bf Given}
a summand term $F(n,s_1,\dots,s_{e})$ which is
hypergeometric\footnote{$F(s)$ is hypergeometric in $s$ iff
$F(s+1)/F(s)=g(s)$ for some fixed rational function $g(s)$.} in
the $s_i$ and in $n$. {\bf Find} a recurrence
\begin{equation}\label{Equ:Recurrence}
p_{\gamma}(n)S(n+\gamma)+\dots+p_0(n)S(n)=0\quad(n\geq0)
\end{equation}
which is satisfied by the hypergeometric multi-sum
$$S(n):=\sum_{s_1}\dots\sum_{s_{e}}F(n,s_1,\dots,s_{e}).$$
In particular, we want to find a recurrence~\eqref{Equ:Recurrence}
which is P-finite, i.e., the coefficients $p_i(n)$ are polynomials
in $n$. Moreover, we suppose that all summations are taken over
finite summand supports. This means, all sums are understood to
extend over all integers, positive and negative, but only finitely
many terms contribute. For example, in $\sum_s\binom{n}{s}$, $n$
a non-negative integer, the summand vanishes if $s<0$ or $s>n$. With this
restriction {\it homogeneous} sum recurrences are guaranteed.

In principle, one could apply the WZ method which is based on
ideas of Sister Celine Fasenmyer and which is described
in~\cite{AequalB}. However, it turns out that all available
implementations of this approach or of variations of it (e.g.,
Wegschaider's algorithm~\cite{Wegschaider:97}) meet in many
applications serious problems of computational complexity. As a
consequence we will follow a different approach which can be
viewed as a simplified variant of Chyzak's
algorithm~\cite{Chyzak:00} within the holonomic system framework~\cite{Zeilberger:90a}.
A full account of computer algebra
details and a comparison to~\cite{Chyzak:00} is given
in~\cite{Schneider:04c}. For further enhancements of this holonomic summation approach in the setting  of difference fields and rings~\cite{Karr:81,Schneider:16}
we refer to~\cite{Schneider:04c,ABRS:12,BRS:18}.
All these new features implemented within the summation package \texttt{Sigma}~\cite{Schneider:07} supported us 
to solve non-trivial problems coming, e.g., from combinatorics~\cite{AndrewsPauleSchneider:04b},
number theory~\cite{SZ:21} or elementary particle physics~\cite{BBFPRRAHSWM:14}.

In this article we will bring in new facets that explain the success of the presented summation method of double and multiple sums.
On one side we will use insight from the summation theory of contiguous relations~\cite{Paule:21} to show the existence of so-called hook-type recurrences which are the basic requirement of our summation approach. Further, we will illustrate in detail how these hook-type recurrences can be utilized to produce without any cost a scalar parameterized recurrence. As a consequence, the entire calculation effort is concentrated in finding a non-trivial rational solution of this derived parameterized recurrence. 
To gain substantial speed ups of our method we present new techniques to compute, e.g., optimal denominator predictions~\cite{Abramov:89a,Abramov:95a,CPS:08} and to discover parts of the the numerator contribution using the Gosper-Petkov{\v s}ek representation~\cite{Petkov:92,AequalB,CPS:08}. All these theoretic and algorithmic contributions will be illustrated by concrete multi-sum examples.

\medskip

The outline of the article is as follows. We start with the base case of our method in Section~\ref{Sec:SingleSums}: the calculation of (hook-type) recurrences of univariate hypergeometric sums. Further algorithmic and theoretic aspects concerning the existence of such recurrences are elaborated in Section~\ref{Sec:ExistenceRec}. Based on this setup, we present our double sum method in 
Section~\ref{Sec:DoubleSums} and supplement it with further examples in Section~\ref{Sec:FurtherExamples}. Furthermore, we explain how this method can be extended to the multi-sum case in Section~\ref{Sec:MultipleSums}. In Section~\ref{Sec:SolveParaRec} we focus on the problem to speed up the key problem of our method. In particular, we focus on various significant improvements to solve parameterized recurrences efficiently. We conclude the article in Section~\ref{Sec:Conclusion}.

\section{Summation Methods for Single Sums}\label{Sec:SingleSums}
Here the basic task is as follows.\\
{\bf Given} a positive integer $\gamma$ and a summand term
$f(n,r)$ which is hypergeometric in $n$ and $r$, {\bf compute} a
P-finite recurrence~\eqref{Equ:Recurrence} which is satisfied by
the sum $S(n):=\sum_rf(n,r)$.

\medskip

In the case that $f(n,r)$ satisfies some mild side conditions this
problem can be solved by applying Zeilberger's
algorithm~\cite{AequalB}. More precisely, one can try to solve the
\notion{creative telescoping problem}: {\bf Find} polynomials
$p_i(n)$, free of $r$, and $g(n,r)$ such that
\begin{equation}\label{CreaEquSingleSumCase1}
p_{\gamma}(n)f(n+\gamma,r)+p_{\gamma-1}(n)f(n+\gamma-1,r)+\dots+p_0(n)f(n,r)=\Delta_rg(n,r);
\end{equation}
$\Delta_r$ denotes the (forward) difference operator defined as
usual by $\Delta_rg(r)=g(r+1)-g(r)$. One can show that if such a
$g(n,r)$ exists, it must be a rational function multiple of
$f(n,r)$. Finally, note that given a solution
for~\eqref{CreaEquSingleSumCase1},
recurrence~\eqref{Equ:Recurrence} is obtained
from~\eqref{CreaEquSingleSumCase1} by summation over all $r$.

\begin{Example}
Within the computer algebra system Mathematica one may use the Paule-Schorn implementation~\cite{PauleSchorn:95} to carry out this summation paradigm. For instance, one can compute for the univariate hypergeometric sum
\begin{equation}\label{eq:PPf1}
	f_1(n,s):=\sum_{k=0}^s\binom{n}{k}^2\binom{n+s-k}{n},
\end{equation}
the recurrence 
\begin{equation}
	-({(1+s)}^2f_1(n,r,s))+(5+6s+2s^2+n+n^2)f_1(n,r,s+1)-{(2+s)}^2f_1(n,r,s+2)=0\label{Rec:Triple:NoRS}
\end{equation}
as follows.
After loading the package

\begin{mma}
	\In <<RISC`fastZeil`\\
	\Print \LoadP{Fast Zeilberger Package\\
		written by Peter Paule, Markus Schorn, and Axel Riese\\
		\copyright\ RISC-JKU}\\
\end{mma}

\noindent into Mathematica and defining its summand $f(n,k,s)$ with

\begin{mma}\MLabel{MMA:summandPPS1}
\In summand = Binomial[n, k]^2 Binomial[n + s - k, n];\\
\end{mma}

\noindent one can solve the creative telescoping problem (here $k$ and $s$ takes over the role of $r$ and $n$ in~\eqref{CreaEquSingleSumCase1})
with the following command:
\begin{mma}
\In Zb[summand, \{k, 0, s\}, s]\\	
\Print If `s' is a natural number and `n' is no negative integer, then:\\
\Out \{-(1+s)^2 SUM[s]+(5+n+n^2+6 s+2 s^2) SUM[1+s]-(2+s)^2 SUM[2+s]==0\} \\
\end{mma}

\bigskip

Alternatively, one may use the \texttt{Sigma}~package~\cite{Schneider:07} 
\begin{mma}
\In \MText{<<Sigma.m} \\
\Print \LoadP{Sigma - A summation package by Carsten Schneider
		\copyright\ RISC-Linz}\\
\end{mma}
\noindent by inserting the input sum
\begin{mma}\MLabel{MMA:f1InputSumSigma}
\In f1 = SigmaSum[
SigmaBinomial[n, k]^2 SigmaBinomial[n + s - k, n], {k, 0, s}]\\
\Out \sum_{k=0}^s \binom{n}{k}^2 \binom{-k
	+n
	+s
}{n}\\
\end{mma}
\noindent and executing the following function call:
\begin{mma}
\In GenerateRecurrence[f1,s]\\
\Out \{(1+s)^2 \text{SUM}[s]
+\big(
-5
-n
-n^2
-6 s
-2 s^2
\big) \text{SUM}[(+s]
+(2+s)^2 \text{SUM}[2+s]
==0\}\\	
\end{mma}
\end{Example}

It can be that for a fixed order $\gamma$ there exists only the
trivial solution, i.e., where all the $p_i(n)$ in~\eqref{CreaEquSingleSumCase1} are $0$. In this
case one has to increase the order $\gamma$ incrementally until a
non-trivial solution is computed. Its existence is guaranteed by
the theory explained in~\cite{AequalB}; see also Section~\ref{Sec:ExistenceRec} below.

\subsection{A slight but important variation}

Many identities involve summands in more than one independent
variable. For instance, instead of the summand $f(n,r)$ consider
the summand $f(m,n,r)$, now hypergeometric in $m,n$ and $r$. For
the following it is important to note that completely analogous
to~\eqref{CreaEquSingleSumCase1} one can compute hook-type recurrences
like
\begin{multline}\label{CreaEquSingleSumCase2}
p_{\gamma}(m,n)f(m+1,n,r)\\
+p_{\gamma-1}(m,n)f(m,n+\gamma-1,r)+\dots+p_0(m,n)f(m,n,r)=\Delta_rg(m,n,r)
\end{multline}
if they exist. This task can be accomplished by a variation
of~\cite{AequalB}. Moreover, the question
whether relations like~\eqref{CreaEquSingleSumCase2} do exist,
will be considered in Section~\ref{Sec:ExistenceRec}.
Summing~\eqref{CreaEquSingleSumCase2} over all $r$ (again assuming
finite summand support) yields
\begin{multline}\label{Equ:SingleSumRecCase2}
p_{\gamma}(m,n)S(m+1,n)
+p_{\gamma-1}(m,n)S(m,n+\gamma-1)+\dots+p_0(m,n)S(m,n)=0\quad(m\geq0)
\end{multline}
with $S(m,n)=\sum_rf(m,n,r)$ and where the $p_i(m,n)$ are
polynomials in $m$ and $n$.


\begin{Example}
The calculation of such hook-type recurrences can be accomplished, e.g., with the Paule-Schorn implementation; see~\cite{Paule:21}.
For instance, given the summand $f(n,s,k)$ of~\eqref{eq:PPf1} defined in~\myIn{\ref{MMA:summandPPS1}} (here $n$, $s$ and $k$ take over the role of $m$, $n$ and $r$ in~\eqref{Equ:SingleSumRecCase2}) one can compute the rational functions $\rho_i(n,s,k)\in\QQ(n,s,k)$ with $f(n,s,k)=\rho_0(n,s,k)f_1(n,s,k)$,
$$f(n,s+1,k)=\rho_1(n,s,k)f_1(n,s,k)\quad\text{and}\quad f(n+1,s,k)=\rho_2(n,s,k)f_1(n,s,k)$$
by executing
\begin{mma}
	\In \{\rho_0, \rho_1, \rho_2\} = FunctionExpand\Big[\newline
	\hspace*{3cm}\big\{summand, (summand /. s \to s + 1),(summand /. n \to n + 1)\big\}\Big/summand\Big]\\
	\Out \{1,\frac{1-k+n+s}{1-k+s},\frac{(1+n) (1-k+n+s))}{(1-k+n)^2}\}\\
\end{mma}
\noindent Then we can extract the hook-type recurrence with the Paule-Schorn implementation by executing the function call 
\begin{mma}\MLabel{MMA:GosperCommand}
	\In Gosper[summand, \{k, 0, s\}, Parameterized \to \{\rho_0, \rho_1, \rho_2\}]\\
	\Out \{Sum[(1+n^2+2 s-2 n s+2 s^2) F_0[k]-2 (1+s)^2 F_1[k]+(1+n)^2 F_2[k],\{k,0,s\}]==\vspace*{0.2cm}\newline
	\hspace*{8cm}\frac{2 (n-s)^2 (1+n-s)^2 Binomial[1+n,s]^2}{(1+n)^2}\}\\
\end{mma}
\noindent More precisely, the output yields
$$\sum_{k=0}^s\big[(1+n^2+2 s-2 n s+2 s^2) F_0[k]-2 (1+s)^2 F_1[k]+(1+n)^2 F_2[k]\Big]=\frac{2 (n-s)^2 (1+n-s)^2}{(1+n)^2}\binom{n+1}{s}^2$$
with $F_0[k]=\rho_0(n,s,k)f_1(n,s,k)=f_1(n,s,k)$, $F_1[k]=\rho_1(n,s,k)f_1(n,s,k)=f_1(n,s+1,k)$ and $F_2[k]=\rho_2(n,s,k)f_1(n,s,k)=f_1(n+1,s,k)$. Then splitting the sum into parts and taking care of the summation ranges produces
\begin{equation}\label{Rec:Triple:NoRSN}
	(1+2s+2s^2-2sn+n^2)f_1(n,s)-2{(1+s)}^2f_1(n,s+1)+{(1+n)}^2f_1(n+1,s)=0;
\end{equation}
this example playing an important role in Example~\ref{Exp:Kratt3} below will be explored further in the next Section~\ref{Sec:ExistenceRec}.

Alternatively, one may use the summation package \texttt{Sigma} by taking the input sum~\myIn{\ref{MMA:f1InputSumSigma}} and executing the command

\begin{mma}
	\In GenerateRecurrence[f1, OneShiftIn \to n]\\	
	\Out \{\big(
	1
	+n^2
	+2 s
	-2 n s
	+2 s^2
	\big) \text{SUM}[s]
	-2 (1+s)^2 \text{SUM}[1+s]
	+(1+n)^2 \text{SUM}[1+n,s]
	==0\}\\
\end{mma}
\end{Example}

\section{Existence of Recurrences}\label{Sec:ExistenceRec}

To discuss, in particular, to guarantee the existence of ``hook-type'' recurrences
of the form as in~\eqref{CreaEquSingleSumCase2}
we make use of the approach described in~\cite{Paule:21}. This approach is based
on a \textit{parameterized} version
of Gosper's algorithm containing 
Zeilberger's creative telescoping
as a special instance.
In~\cite{Paule:21} this idea
is used to derive contiguous relations from
telescoping contiguous relations, thus
covering the existence of both
the Zeilberger-type recurrences as in~\eqref{CreaEquSingleSumCase1}
and the hook-type recurrences
as in~\eqref{CreaEquSingleSumCase2}.

As a concrete illustrating example we choose the
hook-type recurrence~\eqref{Rec:Triple:NoRSN} for the sum \eqref{eq:PPf1}
which will play an important role in Example~\ref{Exp:Kratt3}.
As in~\cite{Paule:21} we use the
notation
\begin{equation}
\label{N3}
{_pF_q}\left( \begin{array}{c}
              a_1, \dots, a_p \\
              b_1, \dots, b_q
             \end{array} ;
         z\right)_k := 
\frac{(a_1)_k \cdots (a_p)_k}{(b_1)_k \cdots (b_q)_k} \frac{z^k}{k!},
\end{equation}
where $(x)_k$ is the shifted factorial 
\begin{equation*}
(x)_k = x(x+1)\cdots (x+k-1) \text{\, if\, } k \geq 1 \text{\, and\, } 
(x)_0=1.
\end{equation*}

\begin{rem}
The motivation for the notation~\eqref{N3} 
and for considering recurrences for such
summands where integer shifts in more
than one parameter are allowed goes 
back to Gau{\ss} who was the first to compile
a table of fifteen classical
contiguous relations; e.g., \cite[7.2]{Gauss},
\begin{equation}
\label{E4.40}
(b-a)\, \xFy{2}{1}{a,b}{c}{z} + a\, \xFy{2}{1}{a+1,b}{c}{z} -
b\, \xFy{2}{1}{a,b+1}{c}{z} =0,
\end{equation}
where
\begin{equation*}
{_2F_1}\left( \begin{array}{c}
              a, b \\
              c
             \end{array} ;
         z\right) = 
\sum_{k=0}^\infty \frac{(a)_k (b)_k}{(c)_k k!} z^k,
\end{equation*}
and where the variables $a,b,c$, and $z$ range over $\CC$ with $|z|<1$ as a condition
for convergence.

In~\cite[Def. 3]{Paule:21} the existence
and derivation of contiguous
relations such as~\eqref{E4.40} is algorithmically explained as limiting cases of \textit{telescoping
contiguous relations}. For example, relation~\eqref{E4.40} is obtained by taking
the limit $n\to \infty$ after summing both sides of
\begin{align}
\label{E4.41}
& c_0\cdot\, \xFyk{2}{1}{a,b}{c}{z} + c_1\cdot\, \xFyk{2}{1}{a+1,b}{c}{z} +
c_2\cdot\, \xFyk{2}{1}{a,b+1}{c}{z}  \nonumber \\
&\hspace*{0.2cm} = \Delta_k\, C(k)\, \xFyk{2}{1}{a,b}{c}{z}, \, \, k\geq 0,
\end{align}
over $k$ from $0$ to $n$.
Theorem~1 in~\cite{Paule:21} predicts the
existence of the $c_j$, $0\leq j \leq 2$, as rational
functions in $\CC(a,b,c,z)$, not all zero,
and of a polynomial $C(x)\in \CC[x]$ with $C(0)=0$ and $\deg C(x)\leq 1$. Moreover, as exemplified
in~\cite[Ex.1, Sec. 6]{Paule:21}, these constituents
can be computed via parameterized creative telescoping:
\[
c_0=b-a, c_1=a, c_2=-b, \text{\, and\, }
C(x)=0.
\]
In view of Zeilberger's creative telescoping
paradigm, telescoping contiguous relations
in which shifts in only one variable occur can 
be called of Zeilberger-type. An example is
\begin{align}
\label{E4.51}
& c_0\cdot\, \xFyk{2}{1}{a,b}{c}{z} + c_1\cdot\, \xFyk{2}{1}{a+1,b}{c}{z} +
c_2\cdot\, \xFyk{2}{1}{a+2,b}{c}{z} \hspace{2cm}  \nonumber \\
&\quad = \Delta_k\, C(k)\, \xFyk{2}{1}{a,b}{c}{z}; 
\end{align}
here Theorem~1 in~\cite{Paule:21} predicts 
$C(0)=0$ and $\deg C(x)\leq 2$. Indeed, by
parameterized telescoping one computes~\cite[eqs. (78) and (79)]{Paule:21},
\begin{equation*}
(c_0, c_1, c_2) = \left( a(a-c+1), 
a( (a-b+1)z-2a-2+c), a(a+1)(1-z) \right)
\text{\, and\, } C(x)=x(x+c-1).
\end{equation*}
We remark that~\eqref{E4.51}, after
summation over $k$ from $0$ to $n$,
in the limit $n\to \infty$ turns into
\begin{align}
\label{E4.50}
&(a+1-c)\, \xFy{2}{1}{a,b}{c}{z} + 
\left( (a+1-b)z -2(a+1)+c \right)\, 
\xFy{2}{1}{a+1,b}{c}{z}    \nonumber \\
&\quad +
(1-z)(a+1)\, \xFy{2}{1}{a+2,b}{c}{z} =0, 
\end{align}
which is the first entry (with $a$ replaced by $a+1$) in the list of fifteen fundamental contiguous relations stated by Gau{\ss}~\cite[7.2]{Gauss}.
In the context of the present article taking
such limits is irrelevant. Nevertheless,
we will exploit the theory for parameterized telescoping relations to guarantee the existence
of hook-type recurrences and also for computing them
algorithmically.
\end{rem}

Back to our illustrating example,
the hook-type relation~\eqref{Rec:Triple:NoRSN}
satisfied by $f_1(n,s)$ as in~\eqref{eq:PPf1}.
It is easily verified that $f_1(n,s)$ can
be rewritten as
\begin{equation}\label{eq:PPF1a}
f_1(n,s)= \binom{n+s}{n} F_1(n,s) 
\end{equation}
with
\begin{equation}\label{eq:PPF1b}
F_1(n,s):=
\sum_{k=0}^s \xFyk{3}{2}{-n,-n,-s}{1,-n-s}{1}
=\xFy{3}{2}{-n,-n,-s}{1,-n-s}{1}.
\end{equation}
The latter equality follows from the fact that
the hypergeometric ${}_3F_2$-series terminates at
$k=s$ owing to the factor $(-s)_k$ in
the $k$th summand and $0\leq s\leq n$.

Using~\eqref{eq:PPF1b} the
hook-type relation~\eqref{Rec:Triple:NoRSN}
rewrites into
\begin{align}\label{eq:PP5hg}
&(1+2s+2s^2-2sn+n^2)F_1(n,s)
+{(1+n)}^2 \frac{\binom{n+s+1}{n+1}}{\binom{n+s}{n}} F_1(n+1,s)
-2{(1+s)}^2 \frac{\binom{n+s+1}{n}}{\binom{n+s}{n}} F_1(n,s+1) \nonumber\\
&=(1+2s+2s^2-2sn+n^2)F_1(n,s)
 +{(1+n)} (1+n+s) F_1(n+1,s)\nonumber \\
&\hspace*{0.8cm}-2{(1+s)} (1+n+s) F_1(n,s+1)
=0.
\end{align}

The shift-structure of the hook-type recurrence~\eqref{eq:PP5hg} leads to
conjecture the existence of a telescoping
contiguous relation with left hand side
\begin{equation}\label{eq:PPF1CRlhs}
c_0\cdot\, \xFyk{3}{2}{a,b,c}{d,e}{1} + c_1\cdot\, \xFyk{3}{2}{a-1,b-1,c}{d,e-1}{1} +
c_2\cdot\, \xFyk{3}{2}{a,b,c-1}{d,e-1}{1}.
\end{equation}
Namely, setting $a=-n, b=-n,c=-s, d=1$, and
$e=-n-s$ the ${}_3F_2(\dots)_k$ expressions
in~\eqref{eq:PPF1CRlhs} from left to right
turn into the
summands of $F_1(n,s)$, $F_1(n+1,s)$, and
$F_1(n,s+1)$.

Indeed, the respective telescoping contiguous
relation is predicted as a special
instance of the following general
theorem where $\KK$ is a suitable field
containing $\QQ$.
\begin{Theorem}[Theorem 1A in~\cite{Paule:21}]\label{thm:PP1A}
 Let 
$a_1,\dots,a_{q+1}$ and $b_1,\dots,b_q$ 
be complex parameters. 
 For $0 \leq l \leq q$ let 
 \[
(\alpha_1^{(l)}, \dots, \alpha_{q+1}^{(l)},
\beta_1^{(l)}, \dots, \beta_q^{(l)})
\]
be pairwise different tuples with non-negative integer entries.
Then there exist $c_0, \dots, c_q$ in $\mathbb{K}$, not all $0$,
and a polynomial $C(x) \in \mathbb{K}[x]$ such that for all $k \geq 0$,
\begin{equation}
\label{N6A}
\sum_{l=0}^q 
c_l\cdot 
{_{q+1}F_q}\left( \begin{array}{c}
              a_1+\alpha_1^{(l)}, \dots, a_{q+1} +\alpha_{q+1}^{(l)}\\
              b_1-\beta_1^{(l)}, \dots, b_q - \beta_q^{(l)}
             \end{array} ;
         1\right)_k = 
\Delta_k\, C(k)\,
{_{q+1}F_q}\left( \begin{array}{c}
              a_1, \dots, a_{q+1} \\
              b_1, \dots, b_q
             \end{array} ;
         1\right)_k.
\end{equation}
Moreover, $C(0)=0$,
and if $C(x) \neq 0$, for the polynomial degree of $C(x)$ one has
\begin{equation}
\label{N6A C deg bound}
\deg C(x) \leq 1 + M\text{\ \ where\ \ }
M:=\max_{0\leq l \leq q}\{\alpha_1^{(l)}+ \dots 
+ \alpha_{q+1}^{(l)}+
\beta_1^{(l)}+\dots +\beta_q^{(l)} \}.
\end{equation}
\end{Theorem}
For our case we have 
\[
(a_1,a_2,a_3)=(a-1,b-1,c-1) \text{\, and\, }
(b_1,b_2)=(d,e),
\]
and with 
\begin{gather*}
(\alpha_1^{(0)}, \alpha_2^{(0)}, \alpha_3^{(0)},
\beta_1^{(0)}, \beta_2^{(0)})=
(1,1,1,0,0),\\
(\alpha_1^{(1)}, \alpha_2^{(1)}, \alpha_3^{(1)},
\beta_1^{(1)}, \beta_2^{(1)})=
(0,0,1,0,1),\\
(\alpha_1^{(2)}, \alpha_2^{(2)}, \alpha_3^{(2)},
\beta_1^{(2)}, \beta_2^{(2)})=
(1,1,0,0,1)
\end{gather*}
 the theorem gives
\begin{align}
\label{eq:PPF1CR}
& c_0\cdot\, \xFyk{3}{2}{a,b,c}{d,e}{1} + c_1\cdot\, \xFyk{3}{2}{a-1,b-1,c}{d,e-1}{1} +
c_2\cdot\, \xFyk{3}{2}{a,b,c-1}{d,e-1}{1}  \nonumber \\
&\hspace*{0.2cm} = \Delta_k\, C(k)\, \xFyk{3}{2}{a-1,b-1,c-1}{d,e}{1}, \, \, k\geq 0,
\end{align}
where $C(x)$ is predicted to be a polynomial
such that $C(0)=0$ and $\deg C(x) \leq 1+M=4$.

The computation of $c_0,c_1,c_2$ and $C(x)$
can be done with the summation package \texttt{Sigma} or, alternatively, with
the Paule-Schorn implementation~\cite{PauleSchorn:95} of Zeilberger's algorithm, both written in Mathematica. Concerning
the latter, in~\cite{Paule:21} the reader finds 
various detailed examples how to do this. For
our concrete case~\eqref{eq:PPF1CR}, the program finds:
\begin{align*}
c_0 &=
-(-1 + a) (-1 + b) (-1 + c)
\\
&\hspace*{0.5cm}(a^2 b + a b^2 + c - a^2 c - a b c - 
   b^2 c - d - a b d + a c d + b c d - a b e - c e + a c e + b c e + 
   d e - c d e),\\
c_1 & =-(-1 + a) (-1 + b) (-1 + c) (a - d) (b - d) (-1 + e),\\  
c_2 & =(-1 + a) (-1 + b) (-1 + c) (-1 + a + b - d) (c - d) (-1 + e),
\end{align*}
and 
\begin{align*}
C(x)& =-x (-1 + d + x) (-1 + e + x) 
\\
&\hspace*{0.5cm}(-2 a b + a^2 b + a b^2 - c + 2 a c - 
   a^2 c + 2 b c - a b c - b^2 c + d - a b d - 2 c d + a c d + b c d \\
   & \hspace*{0.5cm}
   + a b x + c x - a c x - b c x - d x + c d x).
\end{align*}
Setting $a=-n, b=-n,c=-s, d=1$, and
$e=-n-s$ results in
\begin{align*}
(c_0,c_1,c_2)
&=\big(
-(1 + n)^3 (1 + s) (1 + n^2 + 2 s - 2 n s + 2 s^2), -(1 + n)^4 (1 + 
   s) (1 + n + s), \\
& \hspace*{0.7cm}
 2 (1 + n)^3 (1 + s)^2 (1 + n + s)
\big),
\end{align*}
and summing~\eqref{eq:PPF1CR} over $k$ from
$0$ to $s+1$ produces~\eqref{eq:PP5hg} which
is equivalent to~\eqref{Rec:Triple:NoRSN}. Note that with the function call~\myIn{\ref{MMA:GosperCommand}} this recurrence has been produced directly with the specialization $a=-n, b=-n,c=-s, d=1$, and
$e=-n-s$.

We conclude this section with a couple of
remarks. First, the theorems from~\cite{Paule:21}
guarantee the existence of hook-type recurrences
for hypergeometric summands of the form as in~\eqref{N3}. 
In addition, the respective telescoping
contiguous relations
can be computed by any implementation of
parameterized telescoping. Finally, we 
remark that recurrences of Zeilberger-type
are covered as a special case. For example,
the relation~\eqref{Rec:Triple:NoRS},
\begin{align}\label{Rec:Triple:Sa}
&-{(1+s)}^2f_1(n,s)+(5+6s+2s^2+n+n^2)f_1(n,s+1)-{(2+s)}^2f_1(n,s+2) \nonumber\\
&=
-{(1+s)}^2 F_1(n,s)
+(5+6s+2s^2+n+n^2) \frac{\binom{n+s+1}{n}}{\binom{n+s}{n}}F_1(n,s+1)\nonumber\\
&\hspace*{0.5cm} -{(2+s)}^2
\frac{\binom{n+s+2}{n}}{\binom{n+s}{n}} F_1(n,s+2)
\nonumber\\
&=
-{(1+s)}^2 F_1(n,s)
+(5+6s+2s^2+n+n^2)\frac{n+s+1}{s+1}F_1(n,s+1)\nonumber\\
&\hspace*{0.5cm}-{(2+s)} \frac{(n+s+1) (n+s+2)}{s+1} F_1(n,s+2)
=0,
\end{align}
is predicted by another special case
of Theorem~\ref{thm:PP1A}. Namely,
\begin{align}
\label{eq:PPF1Z}
& c_0\cdot\, \xFyk{3}{2}{a,b,c}{d,e}{1} + c_1\cdot\, \xFyk{3}{2}{a,b,c-1}{d,e-1}{1} +
c_2\cdot\, \xFyk{3}{2}{a,b,c-2}{d,e-2}{1}  \nonumber \\
&\hspace*{0.2cm} = \Delta_k\, C(k)\, \xFyk{3}{2}{a,b,c-2}{d,e}{1}, \, \, k\geq 0,
\end{align}
for which Theorem ~\ref{thm:PP1A} predicts
for the polynomial $C(x)$ that
$C(0)=0$ and $\deg C(x) \leq 1+M=5$.
With parameterized telescoping one finds:
\begin{align*}
c_0 &=
(-2 + c) (-1 + c) (1 + a - e) (1 + b - e),
\\
c_1 &=
-(-2 + c) (-1 + e) (3 + a + b + a b - 3 c - a c - b c + 2 d - 2 e + 
   2 c e - d e),\\  
c_2 & =(-2 + c) (-1 + c - d) (-2 + e) (-1 + e),
\end{align*}
and 
\[
C(x) =(c - e) x (-1 + d + x) (-1 + e + x).
\]

\bigskip

Note that besides providing a bound on
the order of Zeilberger-type and hook-type
recurrences (and, in general, of
recurrences stemming
from contiguous relations
with arbitray shift pattern)
such kind of prediction also includes a bound on
the degree of the polynomial $C(x)$ in
the $\Delta_k$ part of the telescoping
contiguous relation.

\section{The Double Sum Method}\label{Sec:DoubleSums}

Here the basic task is as follows.\\
{\bf Given} a summand $F(n,r,s)$ which is hypergeometric in $n,r$
and $s$, {\bf compute} a P-finite
recurrence~\eqref{Equ:Recurrence} which is satisfied by the sum
$S(n):=\sum_r\sum_sF(n,r,s)$.

\begin{Example}\label{Exp:StrehlAllInOne}
With our method under consideration we can solve the following
problem. {Given} the double sum
\begin{equation}\label{Equ:StrehlSumTogether}
S(n)=\sum_{r=0}^n\sum_{s=0}^r\binom{n}{r}\binom{n+r}{r}\binom{k}{s}^3,
\end{equation}
{find} a recurrence of the form~\eqref{Equ:Recurrence} with
$\gamma=2$.
\end{Example}

The overall goal of the method is to compute a recurrence of
type~\eqref{CreaEquSingleSumCase1} where $f(n,r)$ is defined to be
the inner sum, i.e.,
$$f(n,r):=\sum_sF(n,r,s).$$
Note that $g(n,r)$ no longer needs to be a rational function
multiple of $f(n,r)$, hence a suitable ansatz for $g(n,r)$ has to
be introduced; see ANSATZ below.
From~\eqref{CreaEquSingleSumCase1} the desired
recurrence~\eqref{Equ:Recurrence} for $S(n)$ is obtained by
summing over all $r$ --- as in Zeilberger's algorithm for single
sums.

In order to find~\eqref{CreaEquSingleSumCase1} we propose the
following method:\\
{\bf First} one computes recurrences of the following form,
\begin{align}
f(n,r+\delta+1)&=\lambda_0(n,r)f(n,r)+\dots+\lambda_{\delta}(n,r)f(n,r+\delta),\label{Equ:RRecurrence}
\intertext{and}
f(n+1,r)&=\mu_0(n,r)f(n,r)+\dots+\mu_{\delta}(n,r)f(n,r+\delta),\label{Equ:RRecurrenceN}
\end{align}
where the $\lambda_i(n,r)$ and $\mu_i(n,r)$ are rational functions
in $n$ and $r$. This can be accomplished by following
Section~\ref{Sec:SingleSums}; the existence is discussed in
Section~\ref{Sec:ExistenceRec}.

\begin{Example}[Cont.]
For $f(n,r)=\sum_{s=0}^r\binom{n}{r}\binom{n+r}{r}\binom{k}{s}^3$
we can compute with~\cite{AequalB} the recurrences
\begin{multline}\label{Exp:Recurrence1}
8(-1+n-r)(n-r)(1+n+r)(2+n+r)f(n,r)\\
+(-1+n-r)(2+n+r)(16+21r+7r^2) f(n,r+1)-{(2+r)}^4f(n,r+2)=0
\end{multline}
and
\begin{equation}\label{Exp:Recurrence2}
(1+n+r)f(n,r)+(-1-n+r)f(n+1,r)=0,
\end{equation}
i.e., we are in the case $\delta=1$. With \texttt{Sigma} (alternatively, one may use the Paule-Schorn implementation), this can be carried out as follows.
\begin{mma}
\In innerSum=\sum_{s=0}^r\binom{n}{r}\binom{n+r}{r}\binom{k}{s}^3;\\
\end{mma}

\begin{mma}\MLabel{MMA:Strehl:recR}
\In recR = GenerateRecurrence[innerSum][[1]] /. SUM \to f\\
\Out 8 (-1+n-r) (n-r) (1+n+r) (2+n+r) f[r]+(-1+n-r) (2+n+r) (16+21 r+7 r^2) f[1+r]-(2+r)^4 f[2+r]==0\\
\end{mma}

\begin{mma}\MLabel{MMA:Strehl:recRN}
\In recRN = GenerateRecurrence[innerSum, 
OneShiftIn \to n][[1]] /. SUM \to f\\
\Out (1+n+r) f[r]+(-1-n+r) f[1+n,r]==0\\
\end{mma}

\end{Example}

\noindent ANSATZ: For $g(n,r)$ one starts with an expression with
undetermined coefficients of the following form,
\begin{equation}\label{DeltaSolutionDoubleSum}
g(n,r)=\phi_0(n,r)f(n,r)+\dots+\phi_{\delta}(n,r)f(n,r+\delta).
\end{equation}
Then the unknown polynomials $p_i(n)$, free of $r$, and the
unknown rational function coefficients $\phi_i(n,r)$ for $g(n,r)$
are computed such that the certificate
recurrence~\eqref{CreaEquSingleSumCase1} holds. In view
of~\eqref{Equ:RRecurrence} and~\eqref{Equ:RRecurrenceN}, the key
observation is that any shift in $n$ and $r$ of $f(n,r)$ and also
$g(n,r)$ can be represented as a linear combination of
$f(n,r),\dots,f(n,r+\delta)$ over rational functions in $n$ and
$r$. Then rewriting both sides of~\eqref{CreaEquSingleSumCase1} in
terms of these generators, allows us to compute the unknown data
by comparing the coefficients of all the $f(n,r+i)$ involved.

More precisely, we proceed as follows. After computing the
recurrences~\eqref{Equ:RRecurrence} and~\eqref{Equ:RRecurrenceN},
in a {\bf second step} we rewrite the right hand side
of~\eqref{CreaEquSingleSumCase1} as a linear combination in
$f(n,r)$, $f(n,r+1)$ up to $f(n,r+\delta)$. Namely, due
to~\eqref{Equ:RRecurrence} and~\eqref{Equ:RRecurrenceN} there
exist rational functions $\psi_i^{(j)}(n,r)$ in $n$ and $r$ such
that for all nonnegative integers $i$,
\begin{equation}
f(n+j,r)=\sum_{i=0}^{\delta}\psi_i^{(j)}(n,r)f(n,r+i).
\end{equation}
Consequently,
\begin{equation}\label{Equ:LHSRewrite}
\sum_{j=0}^{\gamma}p_j(n)f(n+j,r)=\sum_{i=0}^{\delta}f(n,r+i)\sum_{j=0}^{\gamma}p_j(n)\psi_i^{(j)}(n,r).
\end{equation}


\begin{Example}[Cont.]
We make the ansatz
\begin{equation}\label{Exp:gAnsatz}
g(n,r)=\phi_0(n,r)f(n,r)+\phi_1(n,r)f(n,r+1)
\end{equation}
with
\begin{equation}\label{Equ:ExampleAnsatz}
p_0(n)f(n,r)+p_1(n)f(n+1,r)+p_2(n)f(n+2,r)=\Delta_r g(n,r).
\end{equation}
Then using~\eqref{Exp:Recurrence2}, i.e., using
$f(n+1,r)=\frac{(n+1+r)}{(n+1-r)}f(n,r)$ we rewrite the left hand
side of~\eqref{Equ:ExampleAnsatz} to
\begin{multline*}
p_0(n)f(n,r)+p_1(n)f(n+1,r)+p_2(n)f(n+2,r)\\
=f(n,r)\Big(p_0(n)+p_1(n)\frac{n+r+1}{n-r+1}
+p_2(n)\frac{(n+r+1)(n+r+2)}{(n-r+1)(n-r+2)}\Big).
\end{multline*}
\end{Example}

\noindent OBSERVATION: To compare coefficients we represent also
$\Delta_rg(n,r)$ as a linear combination in $f(n,r)$, $f(n,r+1)$
up to $f(n,r+\delta)$. We get
\begin{multline*}
\Delta_rg(n,r)\stackrel{\eqref{DeltaSolutionDoubleSum}}{=}\sum_{i=0}^{\delta}\phi_i(n,r+1)f(n,r+i+1)
-\sum_{i=0}^{\delta}\phi_i(n,r)f(n,r+i)\\
\stackrel{\eqref{Equ:RRecurrence}}{=}\sum_{i=0}^{\delta-1}\phi_i(n,r+1)f(n,r+i+1)
+\phi_{\delta}(n,r+1)\sum_{i=0}^{\delta}\lambda_i(n,r)f(n,r+i)
-\sum_{i=0}^{\delta}\phi_i(n,r)f(n,r+i)\\
=\sum_{i=1}^{\delta}\big(\phi_{i-1}(n,r+1)+\phi_{\delta}(n,r+1)\lambda_i(n,r)-\phi_i(n,r)\big)f(n,r+i)\\
+\big(\phi_{\delta}(n,r+1)\lambda_0(n,r)-\phi_0(n,r)\big)f(n,r).
\end{multline*}
Comparing the coefficients of the $f(n,r+i)$ to those
in~\eqref{Equ:LHSRewrite} results in the coupled system
\begin{align}
\phi_0(n,r)&=\lambda_0(n,r)\phi_{\delta}(n,r+1)-\sum_{j=0}^{\gamma}p_j(n)\psi_0^{(j)}(n,r)\label{Equ:Base}
\intertext{and}
\phi_i(n,r)&=\phi_{i-1}(n,r+1)+\lambda_i(n,r)\phi_{\delta}(n,r+1)-\sum_{j=0}^{\gamma}p_j(n)\psi_i^{(j)}(n,r)\label{Equ:Successive}
\end{align}
for $1\leq i\leq\delta$; see~\cite[Lemma~1]{Schneider:04c}. This
system can be uncoupled by simple linear algebra, i.e., after
triangularization we arrive at the equivalent system consisting of
the equations~\eqref{Equ:Base}, \eqref{Equ:Successive} for $1\leq
i<\delta$ and
\begin{align}\label{Equ:Triangularized}
-\phi_{\delta}(n,r)+\sum_{j=0}^{\delta}\lambda_j(n,r+\delta-j)\phi_{\delta}(n,r+\delta+1-j)=
\sum_{j=0}^{\gamma}p_j(n)\sum_{i=0}^{\delta}\psi_i^{(j)}(n,r+\delta-i);
\end{align}
see~\cite[Lemma~2]{Schneider:04c}. Summarizing, any solution
$\phi_i(n,r)$ and $p_i(n)$ with~\eqref{Equ:Base},
\eqref{Equ:Successive} for $1\leq i<\delta$
and~\eqref{Equ:Triangularized} gives a solution $g(n,k)$
with~\eqref{DeltaSolutionDoubleSum} and $p_i(n)$
for~\eqref{CreaEquSingleSumCase1}.

\begin{Example}[Cont.]
After rewriting $\Delta_rg(n,r)$ as explained above we can
express~\eqref{Equ:ExampleAnsatz} in the form

\begin{multline}
f(n,r)\Big(\phi_1(n,r+1)\frac{8(-1+n-r)(n-r)(1+n+r)(2+n+r)}{{(2+r)}^4}-\phi_0(n,r)\Big)\\
+f(n,r+1)\Big(\phi_0(n,r+1)+\phi_1(n,r+1)
\frac{(-1+n-r)(2+n+r)(16+21r+7r^2)}{{(2+r)}^4}-\phi_1(n,r)\Big)\\
=f(n,r)\Big(p_0(n)+p_1(n)\frac{n+r+1}{n-r+1}+p_2(n)\frac{(n+r+1)(n+r+2)}{(n-r+1)(n-r+2)}\Big).
\end{multline}
By coefficient comparison of the $f(n,r)$ and $f(n,r+1)$ we get
the coupled system
\begin{multline}\label{Exp:Coupled1}
\phi_1(n,r+1)\frac{8(-1+n-r)(n-r)(1+n+r)(2+n+r)}{{(2+r)}^4}-\phi_0(n,r)\\
=p_0(n)+p_1(n)\frac{n+r+1}{n-r+1}+p_2(n)\frac{(n+r+1)(n+r+2)}{(n-r+1)(n-r+2)}
\end{multline}
and
\begin{equation}\label{Exp:Coupled2}
\phi_0(n,r+1)=-\phi_1(n,r+1)
\frac{(-1+n-r)(2+n+r)(16+21r+7r^2)}{{(2+r)}^4}+\phi_1(n,r).
\end{equation}
Finally, shifting~\eqref{Exp:Coupled1} in $r$ and replacing
$\phi_0(n,r+1)$ with~\eqref{Exp:Coupled2} gives
\begin{multline}\label{Exp:Uncoupled}
\frac{8(1-n+r)(2-n+r)(2+n+r)(3+n+r)}{{(3+r)}^4}\phi_1(n,r+2)\\
-(\frac{(1-n+r)(2+n+r)(16+21r+7r^2)}{{(2+r)}^4})\phi_1(n,r+1)-\phi_1(n,r)\\
=p_0(n)+p_1(n)\frac{2+n+r}{n-r}+p_2(n)\frac{(2+n+r)(3+n+r)}{(n-r)(1+n-r)}.
\end{multline}
This means that any solution $\phi_i(n,r)$ and $p_i(n)$
with~\eqref{Exp:Uncoupled} and~\eqref{Exp:Coupled2} gives a
solution for~\eqref{Equ:ExampleAnsatz}. Note that the previous
transformation steps have been carried out only for illustrative
purpose; the equations~\eqref{Exp:Uncoupled}
and~\eqref{Exp:Coupled2} can be obtained directly from the
explicit formulas~\eqref{Equ:Base}, \eqref{Equ:Successive},
and~\eqref{Equ:Triangularized}.
\end{Example}

Hence, in our {\bf third step} we go on as follows. By using the
algorithms given in Section~\ref{Sec:SolveParaRec} we try to find
a rational function $\phi_{\delta}(n,r)$ in $n$ and $r$ and
polynomials $p_j(n)$ such that~\eqref{Equ:Triangularized} holds.
If we succeed in this task, then we can compute $\phi_0(r)$
by~\eqref{Equ:Base}. Finally, by successive application
of~\eqref{Equ:Successive}, we compute the remaining $\phi_i(r)$.

\begin{Example}[Cont.]\label{Exp:Strehl:PDE:Simple}
We apply the algorithm given in Section~\ref{Sec:SolveParaRec}
and compute the solution
\begin{multline}\label{Equ:Paule:AllInOnePh1Sol}
p_0(n)={(1+n)}^3,\quad p_1(n)=(-3-2n)(39+51n+17n^2),\quad
p_2(n)={(2+n)}^3,\quad\text{and}\\
\phi_1(n,r)=-2(3+2n){(1+r)}^4\Big/\big((n-r)(1+n-r)\big)
\end{multline}
for~\eqref{Exp:Uncoupled}; see
Example~\ref{Exp:LA:StrehlTogether}. Together
with~\eqref{Equ:Base} we obtain the solution
\begin{multline}\label{Equ:gSolution}
	g(n,r)=\Big(2((3+2n)(n-r)(4+6n+2n^2+16r+21nr+7n^2r+19r^2+21nr^2+7n^2r^2-8r^4)f(n,r)\\
	-(3+2n)(2+n-r){(1+r)}^4f(1+r))\Big)\Big/\big((n-r)(1+n-r)(2+n-r)\big)
\end{multline}
for~\eqref{Equ:ExampleAnsatz}.  Using \texttt{Sigma} this result can be obtained with the following function calls.

\begin{mma}
\In mySum=\sum_{r=0}^n f[r];\\
\end{mma}

\begin{mma}
\In CreativeTelescoping[mySum, n, \{\{recR, f[r]\}\}, recRN]\\
\Out\big\{\big\{\frac{(1+n)^3}{3+2 n},-39-51 n-17 n^2,\frac{(2+n)^3}{3+2 n},\newline
\hspace*{-0.4cm} \frac{2\big(
	4
	+6 n
	+2 n^2
	+16 r
	+21 n r
	+7 n^2 r
	+19 r^2
	+21 n r^2
	+7 n^2 r^2
	-8 r^4
	\big)}{(1
	+n
	-r
	) (2
	+n
	-r
	)}f[r]
-\frac{2 \big(
	1+4 r+6 r^2+4 r^3+r^4\big)}{(n
	-r
	) (1
	+n
	-r
	)}f[1+r]
\big\}\big\}\\	
\end{mma}

We note that the correctness of the summand recurrence~\eqref{Equ:ExampleAnsatz}
follows by the derivation given above. Namely,
the solution~\eqref{Equ:Paule:AllInOnePh1Sol} and~\eqref{Equ:gSolution} for~\eqref{Equ:ExampleAnsatz} can be verified by 
simply plugging the solution~\eqref{Equ:Paule:AllInOnePh1Sol} into~\eqref{Exp:Uncoupled} and verifies correctness by rational function arithmetic. If one does not trust this derivation, one may repeat the rewrite rules to get the coupled system~\eqref{Equ:Base} and~\eqref{Equ:Successive} and to verify that the computed $\phi_0$, $\phi_1$ and $\phi_2$ are indeed a solution.
To this end, we can compute the
recurrence
\begin{equation}\label{Equ:Rec:AperySchmidtStrehl}
	(n+1)^3S(n+2)-(2n+3)(17n^2+51n+39)S(n+1)+(n+2)^3S(n)=0
\end{equation}
by summing the equation~\eqref{Equ:ExampleAnsatz} with the explicitly given expressions~\eqref{Equ:Paule:AllInOnePh1Sol} and ~\eqref{Equ:gSolution} over given summation range.  Most of these steps can be carried out automatically with \texttt{Sigma} by executing the following command.
\begin{mma}\MLabel{MMA:Strehl:recOut}
\In GenerateRecurrence[mySum, n, \{\{recR, f[r]\}\}, recRN]\\
\Out \{(1+n)^3 SUM[n]-(3+2 n) (39+51 n+17 n^2) SUM[1+n]+(2+n)^3 SUM[2+n]==-4 (3+2 n) f[0]+\frac{2 (3+2 n)}{n (1+n)} f[1]\}\\
\end{mma}

\noindent 

\noindent Finally, we use the knowledge that $f[0]=f(n,0)=1$ and $f[1]=f(n,1)=2n(n+1)$ holds which shows that the right hand side reduces to $0$.
In short we computed (together with a proof) the recurrence~\eqref{Equ:Rec:AperySchmidtStrehl} for the left hand side of the
Ap\'{e}ry--Schmidt--Strehl identity~\cite{Strehl:94}
$$\sum_{r=0}^n\sum_{s=0}^r\binom{n}{r}\binom{n+r}{r}\binom{k}{s}^3=\sum_{r=0}^n\binom{n}{r}^2\binom{n+r}{r}^2.$$
In total we needed 1.7 seconds on a standard notebook (see Footnote~\ref{FootnoteSpec}) to produce this recurrence; more precisely, it took 1.1 seconds to get the recurrences~\myOut{\ref{MMA:Strehl:recR}} and~\myOut{\ref{MMA:Strehl:recRN}} and 0.6 seconds to obtain~\myOut{\ref{MMA:Strehl:recOut}}.
Lastly, using the Zeilberger's algorithm or \texttt{Sigma} we can compute (again together with proof certificates) the same recurrence~\eqref{Equ:Rec:AperySchmidtStrehl}. Finally, checking two initial values proves the identity.
\end{Example}

\noindent SUMMARY: If we succeed in all our three steps, we manage
to compute polynomials $p(n)$, free of $r$, and $g(n,r)$
with~\eqref{DeltaSolutionDoubleSum} such
that~\eqref{CreaEquSingleSumCase1} holds. By telescoping we arrive
at~\eqref{Equ:Recurrence}. Summarizing, the steps of our algorithm
are as follows.

\begin{alg}{\label{Alg:DoubleSum}}{Creative telescoping for hypergeometric double sums.}
{A summand $F(n,r,s)$ which is hypergeometric in $n$, $r$, and
$s$; in addition\footnotemark $\gamma\in\ZZ_{\geq1}$.}
{A recurrence of the form~\eqref{Equ:Recurrence} for the sum
$S(n)=\sum_r\sum_s F(n,r,s)$.}

\item Compute recurrences of the form~\eqref{Equ:RRecurrence}
and~\eqref{Equ:RRecurrenceN} for the sum $f(n,r):=\sum_sF(n,r,s)$
by parameterized creative telescoping: 
Zeilberger's algorithm and its extension for hook-type recurrences. If not possible, output the
comment ``Failure''.

\item Based on~\eqref{Equ:RRecurrence}
and~\eqref{Equ:RRecurrenceN}, compute rational functions
$\psi_i^{(j)}(n,r)$ to set up the linear system consisting of the
equations~\eqref{Equ:Base}, \eqref{Equ:Successive} for $1\leq
i<\delta$, and~\eqref{Equ:Triangularized}.

\item Try to find a rational function $\phi_{\delta}(n,r)$ in $n$
and $r$ and polynomials $p_j(n)$, free of $r$,
with~\eqref{Equ:Triangularized}; see
Section~\ref{Sec:SolveParaRec}. If not possible, output the
comment ``Failure''.

\item Given $\phi_{\delta}(n,r)$, compute the remaining
$\phi_i(n,r)$ by using~\eqref{Equ:Base}
and~\eqref{Equ:Successive}.

\item Take $g(n,r)$ according to~\eqref{DeltaSolutionDoubleSum},
and sum~\eqref{CreaEquSingleSumCase1} over all $r$. RETURN the
resulting recurrence~\eqref{Equ:Recurrence} for $S(n)$.
\end{alg}
\footnotetext{One may also consider the special case $\gamma=0$. In this case the creative telescoping problem reduces to the telescoping problem; for an example in the context of double sums we refer to Section~\ref{Sec:Blodgett--Andrews--Paule Sum}.}

\subsection{More Flexibility in Specifying
Hypergeometric Double
Sums}\label{Sec:RefinedMethod}

The goal is to compute a recurrence of
type~\eqref{CreaEquSingleSumCase1} where for the summand
$f(n,r)=h(n,r)f'(n,r)$ the following property holds. $h(n,r)$ is an expression (e.g., given as a product of binomial coefficients, factorials and Pochhammer symbols) that is
hypergeometric in $n$ and $r$ and $f'(n,r)$ is defined to be
the inner sum, i.e.,
$$f'(n,r):=\sum_sF(n,r,s).$$

\begin{Example}
We rewrite the double sum given in
Example~\ref{Exp:StrehlAllInOne} to
\begin{equation}\label{Equ:StrehlSumPullOut}
S(n)=\sum_{r=0}^n\binom{n}{r}\binom{n+r}{r}
\sum_{s=0}^r\binom{k}{s}^3,
\end{equation}
i.e., we have $f(n,k)=h(n,k)f'(n,k)$ with
$h(n,r)=\binom{n}{r}\binom{n+r}{r}$ and
$f'(n,r)=\sum_{s=0}^r\binom{k}{s}^3$. With our refined method we
can compute a recurrence of the type~\eqref{CreaEquSingleSumCase1}
with $\gamma=2$.
\end{Example}

In order to find~\eqref{CreaEquSingleSumCase1} we propose a
refined version of the method described above.\\
{\bf First} one computes, as above, recurrences of the following
form,
\begin{align}
f'(n,r+\delta+1)&=\lambda_0(n,r)f'(n,r)+\dots+\lambda_{\delta}(n,r)f'(n,r+\delta),\label{Equ:RRecurrenceRef}
\intertext{and}
f'(n+1,r)&=\mu_0(n,r)f'(n,r)+\dots+\mu_{\delta}(n,r)f'(n,r+\delta),\label{Equ:RRecurrenceNRef}
\end{align}
where the $\lambda_i(n,r)$ and $\mu_i(n,r)$ are rational functions
in $n$ and $r$. Moreover, since $h(n,r)$ is a hypergeometric term
in $n$ and $r$, we can compute rational functions $\rho_{i}(n,r)$
and $\nu_i(n,r)$ such that
\begin{equation}\label{Equ:HypTermRel}
h(n,r+i)=\rho_{i}(n,r)h(n,r)\quad\text{and}\quad
h(n+i,r)=\nu_{i}(n,r)h(n,r)
\end{equation}
for $i\geq0$.

\begin{Example}[Cont.]
With~\cite{AequalB} we compute the recurrences
\begin{equation}\label{Exp:Recurrence1Ref}
8{(1+r)}^2f'(n,r)+(16+21r+7r^2)f'(n,r+1)-{(2+r)}^2f'(n,r+2)=0
\end{equation}
and
\begin{equation}\label{Exp:Recurrence2Ref}
f'(n+1,r)-f'(n,r)=0;
\end{equation}
i.e., we are in the case $\delta=1$. Moreover we have
\begin{align*}
\rho_0&=1,&\rho_1&=\frac{(n-r)(n+r+1)}{(r+1)^2},&\rho_2&=\frac{(n-r-1)(n-r)(n-r+1)(n-r+2)}{(r+1)^2(r+2)^2},\\
\nu_0&=1,&\nu_1&=\frac{n+r+1}{n-r+1},&\nu_2&=\frac{(n+r+1)(n+r+2)}{(n-r+1)(n-r+2)}
\end{align*}
for the relations~\eqref{Equ:HypTermRel}.
\end{Example}

Now we follow the same ideas as above. Namely, due
to~\eqref{Equ:RRecurrenceRef}, \eqref{Equ:RRecurrenceNRef}
and~\eqref{Equ:HypTermRel} one can compute rational functions
$\psi_i^{(j)}(n,r)$ in $n$ and $r$ such that for all nonnegative
integers $i$,
\begin{equation}\label{Equ:PsiDetermination}
f(n+j,r)=h(n+j,r)f'(n+j,r)=h(n,r)\sum_{i=0}^{\delta}\psi_i^{(j)}(n,r)f'(n,r+i).
\end{equation}
To this end, one looks for polynomials $p_i(n)$, free of $r$, and
rational function coefficients $\phi_i(n,r)$ such that with
\begin{equation}\label{DeltaSolutionDoubleSumRef}
g(n,r)=h(n,r)\Big(\phi_0(n,r)f'(n,r)+\dots+\phi_{\delta}(n,r)f'(n,r+\delta)\Big)
\end{equation}
the certificate recurrence~\eqref{CreaEquSingleSumCase1} holds.
More precisely, we look for $p_i(n)$ and $\phi_i(n,r)$ such that
the relations
\begin{align}
\phi_0(n,r)&=\lambda_0(n,r)\rho_1(n,r)\phi_{\delta}(n,r+1)-\sum_{j=0}^{\gamma}p_j(n)\psi_0^{(j)}(n,r),\label{Equ:BaseRef}\\
\phi_i(n,r)&=\rho_1(n,r)\phi_{i-1}(n,r+1)+\rho_1(n,r)\lambda_i(n,r)\phi_{\delta}(n,r+1)-\sum_{j=0}^{\gamma}p_j(n)\psi_i^{(j)}(n,r)\label{Equ:SuccessiveRef}
\end{align}
for $1\leq i<\delta$, and
\begin{multline}\label{Equ:TriangularizedRef}
-\phi_{\delta}(n,r)+\sum_{j=0}^{\delta}\lambda_j(n,r+\delta-j)\rho_{\delta+1-j}(n,r)\phi_{\delta}(n,r+\delta+1-j)\\
=\sum_{j=0}^{\gamma}p_j(n)\sum_{i=0}^{\delta}\rho_{\delta-i}(n,r)\psi_i^{(j)}(n,r+\delta-i).
\end{multline}
hold.

\begin{Example}[Cont.]\label{Exp:Strehl:PDE:Split}
The $\psi(n,r)$ in~\eqref{Equ:PsiDetermination} are given by
$\psi_i(n,r):=\nu_i(n,r)$. Hence~\eqref{Equ:TriangularizedRef}
reads as
\begin{multline}\label{Equ:Paule:SplitSol}
\frac{8(-1+n-r)(n-r)(1+n+r)(2+n+r)}{{(1+r)}^2{(3+r)}^2}\phi_1(n,r+2)\\
+\frac{(n-r)(1+n+r)(16+21r+7r^2)}{{(1+r)}^2{(2+r)}^2}\phi_1(n,r+1)-\phi_1(n,r)
=p_0(n)\frac{(n-r)(1+n+r)}{{(1+r)}^2}\\
+p_1(n)\frac{(1+n+r)(2+n+r)}{{(1+r
)}^2}+p_2(n)\frac{(1+n+r)(2+n+r)(3+n+r)}{(1+n-r){(1+r)}^2}.
\end{multline}
Applying the algorithm given in Section~\ref{Sec:SolveParaRec} we
compute the solution
\begin{multline}\label{Exp:PDESolSplit}
p_0(n)={(1+n)}^3,\quad p_1(n)=(-3-2n)(39+51n+17n^2),\quad
p_2(n)={(2+n)}^3,\quad\text{and}\\
\phi_1(n,r)=\frac{2(3+2n){(1+r)}^2(1+n+r)}{1+n-r};
\end{multline}
see Example~\ref{Exp:LA:StrehlPull}. 
Using~\eqref{Equ:BaseRef} we
compute
$$\phi_0(n,r)=\frac{-2(3+2n)(4+6n+2n^2+16r+21nr+7n^2r+19r^2+21nr^2+7n^2r^2-8r^4
)}{(1+n-r)(2+n-r)}.$$ Altogether we obtained the solution $p_i(n)$
and
$$g(n,r)=\binom{n}{r}\binom{n+r}{r}\big(\phi_0(n,r)f'(n,r)+\phi_1(n,r)f'(n,r+1)\big)$$
for~\eqref{CreaEquSingleSumCase1} with~$\delta=1$ and $\gamma=2$.
\end{Example}

We note that the found result~\eqref{Exp:PDESolSplit} is slightly simpler than the one found in~\eqref{Equ:Paule:AllInOnePh1Sol}, i.e., it contains two factors less. In short, one has to reconstruct two factors less to find a solution which means that the underlying problem to solve a linear recurrence gets simpler. This observation will be further explored in Section~\ref{Sec:Append:Preprocessing} below for more involved examples.

%
%
%
%
%
%

\section{Further Examples}\label{Sec:FurtherExamples}

%
%
%
%
%
%
%
%
%

\subsection{Blodgett--Andrews--Paule Sum}\label{Sec:Blodgett--Andrews--Paule Sum}
We prove the identity
\begin{equation}\label{Id:AndrewsPaule}
\sum_{r=0}^n\sum_{s=0}^n\binom{r+s}{r}^2\binom{4n-2r-2s}{2n-2r}=(2n+1)\binom{2n}{n}^2
\end{equation}
from~\cite{AndrewsPaule:93}. Define
$f(n,r):=\sum_{s=0}^n\binom{r+s}{r}^2\binom{4n-2r-2s}{2n-2r}$.
Then by using \texttt{Sigma} or the Paule-Schorn implementation~\cite{PauleSchorn:95} we compute
\begin{multline}\label{Equ:APRecurr}
(n-r)(1+r)(1-2n+2r)f(n,r)+(18+11n+30n^2+32r-4nr+20n^2r+22r^2-
8nr^2+6r^3)f(n,r+1)\\
-(2+r)(27+9n+18n^2+23r-4nr+6r^2)f(n,r+2)+2(2+r){(3+r)}^2f(n,r+3)=0
\end{multline}
which holds for $0\leq r\leq n-3$. Next, we compute with our
double sum method
\begin{multline*}
g(n,r)=\frac{1}{2{(1+2n)}^2}\Big[(-2-r-3r^2-2r^3-2n^2(5+r)-n(7-5r-4r^2))f(n,r)\\
+(1+r)((10+18n^2+n(13-4r)+9r+4r^2)f(n,r+1)-2{(2+r)}^2f(n,r+2))\Big]
\end{multline*}
such that
\begin{equation}\label{Equ:APTelescoping}
\Delta_rg(n,r)=f(n,r)
\end{equation}
holds for $0\leq r\leq n-3$. This implies that
$$\sum_{r=0}^{n-3}f(n,r)=g(n,n-2)-g(n,0).$$
Using Gosper's algorithm~\cite{Gosper:78} (i.e., the Paule-Schorn implementation) or \texttt{Sigma} we obtain $g(n,0)=0$ and
$g(n,n-2)+f(n,n-2)+f(n,n-1)+f(n,n)=(2n+1)\binom{2n}{n}^2$ which
proves~\eqref{Id:AndrewsPaule}. In total we needed 2.5 seconds to establish this identity.

\medskip

\noindent {\it Remark:} Note that the
recurrence~\eqref{Equ:APRecurr} does not hold for $n-2\leq r\leq
n$. Hence we are not allowed to sum~\eqref{Equ:APTelescoping} over
$0\leq r\leq n$; summing over the whole range would give the wrong
result that the left hand side of~\eqref{Id:AndrewsPaule} equals
to $0$.

\subsection{Ahlgren--Rivoal--Krattenthaler--Sum}\label{Sec:Kratt2}
We prove the identity
\begin{equation}\label{Identity:ARK}
\sum_{r=0}^n\binom{n}{r}^2\binom{2n-r}{n}\sum_{s=0}^r\binom{n}{s}^2\binom{n+r-s}{n}=\sum_{r=0}^n(1-7rH_r+7rH_{n-r})\binom{n}{r}^7
\end{equation}
from~\cite{Krattenthaler:04} which extends the family of
identities from~\cite{PauleSchneider:03}. Define
$$f'(n,r):=\sum_{s=0}^r\binom{n}{s}^2\binom{n+r-s}{n},$$
$h(n,r):=\binom{n}{r}^2\binom{2n-r}{n}$, $f(n,r):=h(n,r)f'(n,r)$,
and $S(n):=\sum_{r=0}^nf(n,r)$. Then by using \texttt{Sigma}, or the Paule-Schorn implementation of Zeilberger's
algorithm~\cite{AequalB} and a variation of it presented in Section~\ref{Sec:ExistenceRec}, we
compute the recurrence relations
\begin{equation}\label{Equ:Rec:Ahlgren7A}
-({(1+r)}^2f'(n,r))+(5+n+n^2+6r+2r^2)f'(n,r+1)-{(2+r)}^2f'(n,r+2)=0
\end{equation}
and
\begin{equation}\label{Equ:Rec:Ahlgren7B}
(1+n^2+2r-2nr+2r^2)f'(n,r)-2{(1+r)}^2f'(n,r+1)+{(1+n)}^2f'(n+1,r)=0
\end{equation}
that hold for all $0\leq r\leq n$. Next we compute the certificate
recurrence
$$\Delta_kg(n,k)=p_0(n,k)f(n,k)+\dots+p_3(n,k)f(n+3,k)$$
given by
\begin{equation}\label{Equ:Kratt2:pi}
\begin{split}
p_0(n,r)&={(1+n)}^4(39+33n+7n^2),\\
p_1(n,r)&=-(56667+199575n+
290457n^2+223446n^3+95773n^4+21675n^5+2023n^6),\\
p_2(n,r)&=-(29445+89733n+111973n^2+73282n^3+26575n^4+5073n^5+399n^6),\\
p_3(n,r)&={(3+n)}^4(13+19n+7n^2),
\end{split}
\end{equation}
and
\begin{equation}\label{Equ:Kratt2:g}
g(n,r)=-\frac{(2n+1-r)(\phi_0(n,r)f(n,r)+\phi_1(n,r)f(n,r+1))}{(n+1)(n+2)(n+1-r)^3(n+2-r)^3(n+3-r)^3}h(n,r)
\end{equation}
where\\
\small
$\phi_0(n,r)=1267n^{16}+n^{15}(35590-13937r)+n^{14}(462869-360690r+67228r^2)+
n^{13}(3700744-4292363r+1601250r^2-194306r^3)+n^{12}(20368825-
31155676r+17403282r^2-4275006r^3+375697r^4)+n^{11}(81899154-
154272523r+114318498r^2-42587312r^3+7684974r^4-498204r^5)+
n^{10}(249131528-552266458r+506714004r^2-
253972267r^3+70678802r^4-9532349r^5+435939r^6)+n^9(585775706-
1477875720r+1602466002r^2-1009881505r^3+385686252r^4-81531645r^5
+7803065r^6-226688r^7)+n^8(1078149331-3015101902r+3728141948r^2-
2822002195r^3+1387276771r^4-410819161r^5+61998191r^6-
3794976r^7+48685r^8)+n^7(1562549948-4739718717r+6484904428r^2-
5688812977r^3+3453596348r^4-1352286608r^5+287672055r^6-
27878506r^7+797964r^8+11193r^9)+n^6(1782583091-
5761209036r+8487438846r^2-8355640898r^3+6074061295r^4-
3045608700r^5+862369998r^6-117912992r^7+5576692r^8+110678r^9-
8232r^{10})+78(61128-257256r+424368r^2-342245r^3+133135r^4-
605158r^5+874931r^6-481375r^7+108900r^8-3013r^9-2506r^{10}+
283r^{11})+n^5(1588987638-5396124481r+8344272508r^2-
8918261604r^3+7554494444r^4-4772804891r^5+1743203551r^6-
316363171r^7+21773661r^8+432467r^9-85561r^{10}+1267r^{11})+n^4(
1088253105-3847125006r+6102147702r^2-6815183791r^3+6526854398r^4-
5179434740r^5+2403086680r^6-558320797r^7+52079159r^8+
813797r^9-366821r^{10}+11517r^{11})+n^3(554906820-
2034079575r+3250687390r^2-3620706197r^3+3753196026r^4-
3786628463r^5+2226983897r^6-648088722r^7+78309002r^8+639830r^9-
831274r^{10}+41326r^{11})+n(44801424-178444188r+286511076r^2-
268017747r^3+231495788r^4-453726129r^5+455141086r^6-202147723r^7+
37738137r^8-523202r^9-704439r^{10}+63933r^{11})+n^2
(198939024-757536768r+1209747438r^2-1270694237r^3+
1316530531r^4-1754875242r^5+1324652652r^6-477128371r^7+
72428076r^8-131397r^9-1051417r^{10}+73167r^{11})$\\
\normalsize and\\
\small
$\phi_1(n,r)={(1+r)}^2(1267n^{14}+n^{13}(34323-13937r)+n^{12}(427279-349287r+
64694r^2)+n^{11}(3239142-3997785r+1495784r^2-166432r^3)+
n^{10}(16702404-27663162r+15686820r^2-3539388r^3+233555r^4)+
n^9(61957608-129089860r+98653180r^2-33877490r^3+4514021r^4-
173215r^5)+n^8(170471516-428918244r+414295520r^2-192602705r^3+
38897286r^4-3007648r^5+55937r^6)+n^7(353346582-1043782832r+
1223781806r^2-722527950r^3+196740999r^4-23008400r^5+870673r^6+
3591r^7)+n^6(554331233-1883343858r+2606958078r^2-1877626092r^3+
646725800r^4-101747168r^5+5872142r^6+36536r^7-6965r^8)+78(
61128-379512r+1061136r^2-1705493r^3+1411049r^4-592120r^5+
110692r^6-366r^7-2789r^8+283r^9)+n^5(654872133-2519285191r+
4035037044r^2-3448656883r^3+1443487563r^4-286571923r^5+
22400799r^6+149302r^7-75311r^8+1267r^9)+n^4(573379725-
2467242453r+4503468974r^2-4476586363r^3+2215350588r^4-
533046699r^5+52847951r^6+308630r^7-337012r^8+11517r^9)+n^3(
360735780-1719341103r+3534592606r^2-4024522064r^3+2308453137r^4-
654819699r^5+78949756r^6+325414r^7-799433r^8+41326r^9)+n(
40033440-228909132r+581493852r^2-840158733r^3+621255632r^4-
231745735r^5+38124358r^6-26621r^7-746298r^8+63933r^9)+n^2(
154137600-807300900r+1851811578r^2-2386528406r^3+1563194205r^4-
512350625r^5+72945156r^6+134868r^7-1060651r^8+73167r^9))$.\\
\normalsize

\noindent This shows that the left hand side
of~\eqref{Identity:ARK} fulfills the recurrence relation
\begin{multline}\label{Equ:Rec:Ahlgren7}
{(1+n)}^4(39+33n+7n^2)S(n)\\-(56667+199575n+
290457n^2+223446n^3+95773n^4+21675n^5+2023n^6)
S(n+1)\\
-(29445+89733n+111973n^2+73282n^3+26575n^4+5073n^5+399n^6)S(n+2)\\
+{(3+n)}^4(13+19n+7n^2)S(n+3)=0.
\end{multline}
The total caluclation time of this recurrence took 4.8 seconds; more precisely, 1.3 seconds for recurrences~\eqref{Equ:Rec:Ahlgren7A} and ~\eqref{Equ:Rec:Ahlgren7B} of the inner sum and 3.8 seconds for the recurrence~\eqref{Equ:Rec:Ahlgren7A} of the double sum.
In~\cite{PauleSchneider:03} the same recurrence relation~\eqref{Equ:Rec:Ahlgren7} has been
derived for the right hand side of~\eqref{Identity:ARK}. Checking
the first three initial values proves~\eqref{Identity:ARK}.

\section{The Method Extended to Multiple Sums}\label{Sec:MultipleSums}

Based on what we said about single and double sums we are in the
position to deal with the general problem stated at the beginning of
Section~\ref{Sec:Introduction}.\\
\noindent {\bf Given} a summand $F(m,n,r,s_1,\dots,s_e)$ which is
hypergeometric in $m,n,r$ and the $s_i$, {\bf compute} a P-finite
recurrence
\begin{equation}\label{Equ:MultipleSumRecCase}
p_{\gamma}(m,n)S(m,n+\gamma)+\dots+p_0(m,n)S(m,n)=0
\end{equation}
(resp.\ a P-finite recurrence~\eqref{Equ:SingleSumRecCase2}) which
is satisfied by the sum
$$S(m,n)=\sum_{r}\sum_{s_1}\dots\sum_{s_e}F(m,n,r,s_1,\dots,s_e).$$

As with double sums the overall goal of the method is to compute a
certificate recurrence of the form
\begin{equation}\label{CreaEquMultipleSumCase}
p_{\gamma}(m,n)f(m,n+\gamma,r)+\dots+p_0(m,n)h(m,n,r)=\Delta_{r}g(m,n,r)
\end{equation}
(resp.~\eqref{CreaEquSingleSumCase2}) where we define $f(m,n,r)$
as
\begin{equation}\label{SummandDoubleSum}
f(m,n,r):=\sum_{s_1}\dots\sum_{s_{e}}F(m,n,r,s_1,\dots,s_{e}),
\end{equation}
and where $g(m,n,r)$ is suitably chosen. Then
from~\eqref{CreaEquMultipleSumCase}
(resp.~\eqref{CreaEquSingleSumCase2}) the desired
recurrence~\eqref{Equ:MultipleSumRecCase}
(resp.~\eqref{Equ:SingleSumRecCase2}) for $S(m,n)$ is obtained by
summation over all $r$.

To find~\eqref{CreaEquMultipleSumCase} we proceed analogously to
the double sum case. Namely, we first try to derive recurrences of
the form
\begin{align}
f(m,n,r+\delta+1)&=\lambda_0(m,n,r)f(m,n,r)+\dots+\lambda_{\delta}(m,n,r)f(m,n,r+\delta),\label{Equ:RRecurrenceG}
\intertext{and}
f(m,n+1,r)&=\mu_0(m,n,r)f(m,n,r)+\dots+\mu_{\delta}(m,n,r)f(m,n,r+\delta),\label{Equ:RRecurrenceNG}
\end{align}
where the $\lambda_i(m,n,r)$ and $\mu_i(m,n,r)$ are rational
functions in $m,n$ and $r$. Afterwards we apply the same method as
in the double sum case in order to compute all the components for
the certificate recurrence~\eqref{CreaEquMultipleSumCase}.

Otherwise, if we look for~\eqref{Equ:SingleSumRecCase2}, we
suppose that we have computed besides~\eqref{Equ:RRecurrenceG}
and~\eqref{Equ:RRecurrenceNG} a hook-type recurrence of the form
\begin{align}\label{Equ:RRecurrenceMG}
f(m+1,n,r)&=\nu_0(m,n,r)f(m,n,r)+\dots+\nu_{\delta}(m,n,r)f(m,n,r+\delta).
\end{align}
Then, we can represent the left hand side
of~\eqref{CreaEquSingleSumCase2} in terms of the generators
$f(m,n,r),...,f(m,n,r+\delta)$. More precisely, as in the double
sum case~\eqref{Equ:LHSRewrite} we can compute rational functions
$\psi_i^{(j)}(m,n,r)$ in $m$, $n$ and $r$ such that
\begin{equation*}
p_{\gamma}(m,n)f(m+1,n,r)+\sum_{j=0}^{\gamma-1}p_j(m,n)f(m,n+j,r)=\sum_{i=0}^{\delta}f(m,n,r+i)\sum_{j=0}^{\gamma}p_j(m,n)\psi_i^{(j)}(m,n,r).
\end{equation*}
holds. Given this representation, we can proceed as in the double
sum case in order to compute all the components for the
certificate recurrence~\eqref{Equ:SingleSumRecCase2}.

Note that our refined method in Subsection~\ref{Sec:RefinedMethod}
can be carried over analogously to the multiple sum case.

Summarizing, in order to apply the above strategy there remains
the task to compute the recurrences of the
type~\eqref{Equ:RRecurrenceG}, \eqref{Equ:RRecurrenceNG}
and~\eqref{Equ:RRecurrenceMG}. This gives rise to the following situations.

\begin{description}
\item[Base case] If $f(n,m,r)$ is a single sum, i.e., $e=1$, we
can apply Zeilberger's algorithm to get~\eqref{Equ:RRecurrenceG},
or a variation of it to obtain~\eqref{Equ:RRecurrenceNG}; see
Section~\ref{Sec:SingleSums}. Similarly, we can
compute~\eqref{Equ:RRecurrenceMG} by a slightly more general
variation; see~Section~\ref{Sec:ExistenceRec}.


\item[Reduction] Otherwise, we apply again the method described in
this section, but this time for a multi-sum reduced by one sum.
This means that by recursion we end up eventually in the base
case.
\end{description}

\begin{Example}\label{Exp:Kratt3}
We illustrate how one can prove identity
\begin{multline}\label{Ahlgren:09}
\quad\sum_{r=0}^n\binom{n}{r}^2\binom{2n-r}{n}\sum_{s=0}^r\binom{n}{s}^2\binom{n+r-s}{n}\sum_{k=0}^s\binom{n}{k}^2\binom{n+s-k}{n}\\
=\sum_{r=0}^n(1-9rH_r+9rH_{n-r})\binom{n}{r}^9 \quad
\end{multline}
from~\cite{Krattenthaler:04}. Define
$$f_1(n,r,s):=\sum_{k=0}^s\binom{n}{k}^2\binom{n+s-k}{n},$$
$h_1(n,r,s):=\binom{n}{s}^2\binom{n+r-s}{n}$,
$f_2(n,r):=\sum_{s=0}^rh_1(n,r,s)f_1(n,r,s)$,
$h_2(n,r):=\binom{n}{r}^2\binom{2n-r}{n}$, and let $S(n)$ be the
left hand side of~\eqref{Ahlgren:09}, i.e.,
$S(n):=\sum_{r=0}^nh_2(n,r)f_2(n,r)$. In order to compute a
recurrence for $S(n)$, we apply the machinery of \texttt{Sigma} or the algorithms~\cite{AequalB}
and~\cite{Paule:21} (see Sections~\ref{Sec:SingleSums} and~\ref{Sec:ExistenceRec}) to obtain the recurrence relations
\begin{gather}
-({(1+s)}^2f_1(n,r,s))+(5+6s+2s^2+n+n^2)f_1(n,r,s+1)-{(2+s)}^2f_1(n,r,s+2)=0,\label{Rec:Triple:S}\\
(1+2s+2s^2-2sn+n^2)f_1(n,r,s)-2{(1+s)}^2f_1(n,r,s+1)+{(1+n)}^2f_1(n+1,r,s)=0\label{Rec:Triple:SN}
\intertext{which are equivalent to~\eqref{Rec:Triple:NoRS} and~\eqref{Rec:Triple:NoRSN} with $f_1(n,r,s)=f_1(n,s)$. Further we get trivially} f_1(n,r,s)-f_1(n,r+1,s)=0.\label{Rec:Triple:SR}
\end{gather}
Given~\eqref{Rec:Triple:S} and~\eqref{Rec:Triple:SR} we apply our
double sum method from Section~\ref{Sec:DoubleSums} to compute
the recurrence relation
\begin{multline}\label{Rec:Triple:R}
{(1+r)}^2{(2+r)}^2f_2(n,r)-{(2+r)}^2(14+3n+3n^2+12r+3r^2)f_2(n,r+1)\\
+(133+n^4+200r+115r^2+30r^3+3r^4-n^3(3+2r)+n^2(13+12r+3r^2)\\
+n(17+14r+3r^2))f_2(n,r+2)-{(3+r)}^4f_2(n,r+3)=0.
\end{multline}
As explained above, we compute in addition the recurrence relation \small
\begin{multline}\label{Rec:Triple:RN}
-({(1+r)}^2(2n^4-n^3(7+10r)+n^2(20+42r+24r^2)-n(15+68r+78r^2+
28r^3)\\
+2(6+24r+40r^2+28r^3+7r^4))f_2(n,r)) +(91+5n^6+450r+971r^2
+1084r^3+659r^4+210r^5+28r^6-3n^5(3+8r)\\
+n^4(29+66r+57r^2)
-n^3(-9+64r+123r^2+70r^3)+n^2(101+210r+246r^2+174r^3+57r^4)\\
-n(54+362r+633r^2+520r^3+222r^4+42r^5))f_2(n,r+1)\\
-{(2+r)}^4(5+5n^2+14r+14r^2-2n(2+7r))\,f_2(n,r+2)=-({(1+n)}^4{(1+r)}^2f_2(n+1,r))
\end{multline}
\normalsize

\noindent by using besides~\eqref{Rec:Triple:S}
and~\eqref{Rec:Triple:SR} the recurrence
relation~\eqref{Rec:Triple:SN}. Given the two
recurrences~\eqref{Rec:Triple:R} and~\eqref{Rec:Triple:RN}, we are
in the position to apply our method again as in the double sum
case. This gives the recurrence

\small
\begin{multline}\label{Equ:RecForTriple}
{(1+n)}^6{(2+n)}^2(126186232584+359847089412n+447038924854n^2+
315988281882n^3\\
+139000794255n^4 +38967288138n^5+6799034214n^6
+675116208n^7+29211759n^8)S(n)\\
+2{(2+n)}^2(9449901867223980+65177937447506574n+206795641058521957n^2\\
+400003560150467208n^3+526934624462960841n^4+
500054178553882862n^5\\
+352526028922986741n^6+187547382614273601n^7+75664907849081395n^8+23037690482849736n^9\\
+5211078007675644n^{10}+849237300832941n^{11}+
94267319550444n^{12}+6380425909278n^{13}\\
+198698384718n^{14})S(n+1)-3(99381765767163760+720338927889449008n+
2427055018593335824n^2\\
+5046939121521308492n^3+
7251199169750148467n^4+7634448497599004444n^5+
6094496182619292815n^6\\
+3763786379996759276n^7+1817742639895041823n^8+688977924255751768n^9+
204313397754918826n^{10}\\
+46914883776289584n^{11}+8179105939324551n^{12}+1046803624503588n^{13}+
92772291582963n^{14}\\
+5087571879456n^{15}+130079962827n^{16})S(n+2)\\
-{(3+n)}^2(1657317485213296+10358247512403136n+29676907405770592n^2\\
+51669502990568780n^3+61088527857001943n^4+51897294744470249n^5
+32681221486607779n^6\\
+15503112379989763n^7+5569174593112480n^8+1508250655288332n^9+
303253251903666n^{10}\\
+43913846933991n^{11}+4331266602147n^{12}+
260552661525n^{13}+7215304473n^{14})S(n+3)\\
+{(3+n)}^2{(4+n)}^6(3576422026+16265263120n+32031965452n^2+
35670510738n^3\\
+24565622625n^4+10714664718n^5+2891150010n^6+441422136n^7+29211759n^8)S(n+4)=0.
\end{multline}
\normalsize

\noindent For the calculation of the recurrences~\eqref{Rec:Triple:S} and~\eqref{Rec:Triple:SN} of the innermost sum we needed 1.3 seconds, for the recurrences of the double sum we used 5.5 seconds and for the final output recurrence~\eqref{Equ:RecForTriple} of the triple sum we used 8.1 seconds. Thus the full calculation could be accomplished in less than 15 seconds.
By applying the summation package \SigmaP,
see~\cite{PauleSchneider:03}, we arrive at the same recurrence for
the right hand side of~\eqref{Ahlgren:09}. Checking the first
initial values proves identity~\eqref{Ahlgren:09}.

\end{Example}

\section{Speeding Up our Multi-Sum Method}\label{Sec:SolveParaRec}

	The computational backbone concerning efficiency of our method is introduced in this section.	
		As elaborated in Sections~\ref{Sec:DoubleSums} and~\ref{Sec:MultipleSums} the creative telescoping problem~\eqref{CreaEquSingleSumCase1} for double sums and more generally multi-sums can be reduced efficiently to the problem to 
		solve parameterized linear recurrences of the form~\eqref{Equ:Triangularized}
		or~\eqref{Equ:TriangularizedRef} by using rewrite rules combing from the linear (hook-type) recurrences of the summand. 
		 In view of~\eqref{Equ:Triangularized}
	and~\eqref{Equ:TriangularizedRef} we consider the following
	problem:\\[0.3cm]
	{\bf Given} a rational function field $\KK(r)$,
	$a_0(r),\dots,a_{\delta}(r)\in\KK[r]$ with $a_0\,a_r\neq0$ and
	$f_0(r),\dots,f_{\gamma}(r)\in\KK[r]$, {\bf find} all solutions
	$c_0,\dots,c_{\gamma}\in\KK$ and $g(r)\in\KK(r)$ of the parameterized recurrence
	\begin{equation}\label{PLDE}
		a_{\delta}(r)\,g(r+\delta)+a_{\delta-1}(r)\,g(r+\delta-1)+\dots+a_0(r)\,g(r)=c_0\,f_0(r)+\dots+c_{\gamma}\,f_{\gamma}(r).
	\end{equation}
	
	\medskip
		
	\noindent In all our examples the 
	full calculation of our proposed summation method, \textit{excluding} the task to find a solution of~\eqref{PLDE}, took at most 2 seconds. In a nutshell, almost all of the calculation time is used to solve the underlying parameterized recurrence.
	
	In the following we introduce the basic mechanism implemented within \texttt{Sigma} and present various improvements that lead to significant speed-ups. For instance, combining all these enhancements finally enabled us to compute recurrences of the double sum and triple sum on the left hand sides of~\eqref{Identity:ARK} and~\eqref{Ahlgren:09}
	in less than 5 and 15 seconds, respectively.
	
	The basic algorithm works as follows.\\[0.2cm]
	\noindent{\bf(1)} In a first step we compute a \notion{denominator
		bound for~\eqref{PLDE}}, i.e., a non-zero polynomial
	$d(r)\in\KK[r]$ such that for any solution $g(r)\in\KK(r)$ and
	$c_i\in\KK$ with~\eqref{PLDE} we have $d(r)\,g(r)\in\KK[r]$; this
	task can be accomplished by Abramov's algorithm
	in~\cite{Abramov:89a}, \cite{Abramov:95a}. Here we use the equivalent compact formula given in~\cite{CPS:08}:
	\begin{equation}\label{Equ:AbramovBound}
		d(r)=\gcd\Big(\prod_{i=0}^Da_0(r+i),\prod_{i=0}^Da_{\delta}(r-\delta-i)\Big)
	\end{equation}
	where $D\in\ZZ\cup\{-\infty\}$ is the dispersion of the coefficients $a_{\delta}(r)$ and $a_{0}(r)$ defined by $$D=\max(h\in\ZZ_{\geq0}\mid\gcd(a_{\delta}(r-\delta),a_0(r+h))=1);$$ for a generalized formula that holds for coupled systems in $\Pi\Sigma$-extensions~\cite{Karr:81,Schneider:01a} we refer to~\cite{MS:18}.  Note that in basically all our applications the polynomials $a_0(r)$ and $a_{\delta}(r)$ are already given in factored form and thus the gcd in~\eqref{Equ:AbramovBound} can be read off. In particular the result $d(r)$ can be also given directly in its factored form, which we will need in~\eqref{Equ:DenBoundFactored} below.\\
	Then, given such a denominator bound $d(r)$, it suffices to look
	for all $g'(r)\in\KK[r]$ and $c_i\in\KK$ with
	\begin{align}\label{Equ:DenFreeLDE}
		\frac{a_{\delta}(r)}{d(r+\delta)}\,g'(r+\delta)+\dots+
		\frac{a_{1}(r)}{d(r+1)}\,g'(r+1)
		+\frac{a_0(r)}{d(r)}\,g'(r)=c_0\,f_0(r)+\dots+c_{\gamma}\,f_{\gamma}(r).
	\end{align}
	Namely, given all such solutions $g'(r)$ and $c_i$, we obtain all
	the solutions of~\eqref{PLDE} with $\frac{g'(r)}{d(r)}$ and
	$c_i$.\\
	\noindent{\bf(2)} The next step consists of bounding the
	polynomial degree of the possible solutions $g'(r)\in\KK[r]$, say
	with $b\in\NN$. In
	\cite{Abramov:89b,Petkov:92,Abramov:95b,AequalB} several
	algorithms are introduced that find such a \notion{degree bound
		$b$ for~\eqref{Equ:DenFreeLDE}}. Note that
	all these algorithms are equivalent; see~\cite{PetkovWeix}.\\
	\noindent{\bf(3)} Finally, substituting the possible solutions
	$g'(r)=g_b\,r^b+g_{b-1}\,r^{b-1}+\dots+g_0$
	into~\eqref{Equ:DenFreeLDE} leads by coefficient comparison to a
	linear system of equations. Solving this system enables one to
	construct all the solutions for~\eqref{Equ:DenFreeLDE} and hence
	for~\eqref{PLDE}. More precisely, one can compute a basis of the $\KK$-vector space
	\begin{equation*}
		V=\{(c_1,\dots,c_{\delta},g)\in\KK^{\delta}\times\KK(r)\mid \text{equation~\eqref{PLDE} holds}\}
	\end{equation*} 
	whose 
	dimension is at most $\delta+1+\gamma$.

	\begin{Example}[Cont.
		Example~\ref{Exp:Strehl:PDE:Simple}]\label{Exp:LA:StrehlTogether}
		Following the algorithm from above we compute
		$\Phi_1(n,r)\in\QQ(n)(r)$ and $p_i(n)\in\QQ(n)$ such
		that~\eqref{Exp:Uncoupled} holds: First we compute the denominator
		bound $d(r)=(n-r)(n+1-r)$ using the formula~\eqref{Equ:AbramovBound}. As a result, \eqref{Equ:DenFreeLDE} reads as
		\small
		\begin{multline}\label{Equ:PolyDFE}
			8(-2+n-r)(-1+n-r)(n-r)(1+n-r){(2+r)}^4(2+
			n+r)(3+n+r)g'(r+2)\\
			+(-1+n-r)(n-r)(1+n-r){(3+r)}^4(
			2+n+r)(16+21r+7{r}^2)g'(r+1)\\
			-((n-r)(1+n-r){(2+r)}^4{(3+r)}^4)g'(r)
			=p_0(n-r)(1+n-r){(2+r)}^4{(3+r)}^4\\
			+p_1(1+n-r){(2+r)}^4{(3+r)}^4(2+n+r)
			+p_2{(2+r)}^4{(3+r)}^4(2+n+r)(3+n+r).
		\end{multline}
		\normalsize Next, we compute the degree bound $b=4$ for the
		polynomial solutions $g'(r)\in\QQ(n)[r]$. Finally, substituting
		the possible solutions $g'(r)=\sum_{i=0}^4g'_ir^i$ and
		$p_i\in\QQ(n)$ into~\eqref{Equ:PolyDFE} leads by coefficient
		comparison to a linear system with $13$~equations in $8$~unknowns
		($p_0,p_1,p_2,g'_0,\dots,g'_4$). Note that this system requires
		$23576$ bytes of memory in the computer algebra system Mathematica
		. To this end, solving this system gives the solution
		$p_0={(1+n)}^3$ $p_1=(-3-2n)(39+51n+17n^2)$, $p_2(n)={(2+n)}^3$,
		and $g'(r)=-2(3+2n){(1+r)}^4$, and hence the
		solution~\eqref{Equ:Paule:AllInOnePh1Sol}
		for~\eqref{Exp:Uncoupled}.
	\end{Example}

	\begin{Example}[Cont. Example~\ref{Exp:Strehl:PDE:Split}]\label{Exp:LA:StrehlPull}
		Completely analogously, we solve~\eqref{Equ:Paule:SplitSol}.
		Namely, we compute the denominator bound $d(r)=n+1-r$, afterwards
		we consider the corresponding problem of the
		form~\eqref{Equ:DenFreeLDE}, compute the degree bound $b=3$, and
		set up a linear system with $10$ equations in $7$ unknowns.
		Solving this system gives the solution~\eqref{Exp:PDESolSplit}
		for~\eqref{Equ:Paule:SplitSol}. Note that in comparison to
		Example~\ref{Exp:LA:StrehlTogether} the degree of the denominator
		bound and hence also the degree bound is reduced by one. This
		leads us to a smaller equation system, namely $10\times3$ instead
		of $13\times8$; in Mathematica we need only $15408$ bytes instead
		of $23576$ bytes to store the system.
	\end{Example}
	
	\subsection{Preprocessing of the input sums}\label{Sec:Append:Preprocessing}
		
	The observation described in Example~\ref{Exp:LA:StrehlPull} holds in all our
	examples. Pulling out expressions from the inner sum,
	like~\eqref{Equ:StrehlSumTogether}
	and~\eqref{Equ:StrehlSumPullOut}, and applying our refined
	summation method from Subsection~\ref{Sec:RefinedMethod} amounts
	to find a solution~\eqref{PLDE} with a smaller degree of the
	denominator. In particular this reduces considerably the size of
	the linear system and the amount of time to find the solutions.
	
	\begin{Example}[Cont.
		Subsection~\ref{Sec:Kratt2}]\label{Exp:SimpleAlg:Kratt2} In order
		to compute~\eqref{Equ:Kratt2:pi} and~\eqref{Equ:Kratt2:g}, we
		apply our method in Subsection~\ref{Sec:RefinedMethod} which
		reduces to a problem of the type~\eqref{PLDE}. In order to solve
		this problem, we compute the denominator bound
		\begin{equation}\label{Equ:Kratt2Den:Pull:No}
			d(r)=(n+1-r)^3(n+2-r)^6(n+3-r)^3
		\end{equation}
		and the degree bound $b=15$. This finally gives a linear system
		with $30$ equations in $20$ unknowns. In Mathematica this system
		requires 0.67~MB of memory. Solving this system can be carried out
		in $5.7$ seconds using $28$~MB memory; compare the first
		row of Table~\ref{T:DoublePull}.\\
		By doing the same computations without pulling out factors from
		the innermost sum, i.e., considering the sum
		\begin{equation}\label{DoubleSumAhlgren7:InOne}
			\sum_{r=0}^n\sum_{s=0}^r\binom{n}{r}^2\binom{2n-r}{n}\binom{n}{s}^2\binom{n+r-s}{n},
		\end{equation}
		we compute the denominator bound
		\begin{equation}\label{Equ:Kratt2Den:Simple:No}
			d(r)=(n-r)^3(n+1-r)^6(n+2-r)^6(n+3-r)^3
		\end{equation}
		and the degree bound $b=21$; for the properties of the linear
		system see the first row of Table~\ref{T:DoubleAll}.
	\end{Example}
	
	\begin{Example}[Cont. Example~\ref{Exp:Kratt3}]
		In order to find the recurrence~\eqref{Equ:RecForTriple} for the
		triple sum $S(n)$, we apply our refined method in
		Subsection~\ref{Sec:RefinedMethod}. We get the denominator bound
		\begin{equation}\label{Equ:Kratt3Den:Pull:No}
			d(r)=(n+1-r)^3(n+2-r)^6(n+3-r)^6(n+4-r)^3(r+1)^2
		\end{equation}
		and the degree bound $b=25$. For the linear system see the first
		row of Table~\ref{T:TriplePull}.\\
		Applying our method without pulling out factors, i.e., considering
		the sum
		\begin{equation}\label{Ahlgren:09AllInOne}
			\quad\sum_{r=0}^n\sum_{s=0}^r\sum_{k=0}^s\binom{n}{s}^2\binom{n+r-s}{n}\binom{n}{r}^2\binom{2n-r}{n}
			\binom{n}{k}^2\binom{n+s-k}{n}
		\end{equation}
		leads us to a denominator bound
		\begin{equation}\label{Equ:Kratt3Den:Simple:No}
			d(r)=(n-1-k)^3(n-r)^6(n+1-r)^9(n+2-r)^9(n+3-r)^6(n+4-r)^3
		\end{equation}
		and a degree bound $b=41$; for the properties of the linear system
		see the first row of Table~\ref{T:TripleAll}.
	\end{Example}
	
	\footnotesize
	\begin{table}[h]
		\caption{\label{T:DoubleAll}Double
			sum~\eqref{DoubleSumAhlgren7:InOne} (without preprocessing)}
		\begin{tabular}{|c|r@{$\times$}l|c|c|c|}\hline
			Improvements&equs&unknowns&size of system&total time\footnotemark&
			total memory$^{\ref{FootnoteSpec}}$\\
			\hline\hline None&38&26&1.1~MB&12.2~s&28~MB\\
			\hline I&26&20&0.36~MB&4.0~s&23~MB\\
			\hline II&25&26&0.65~MB&5.8~s&26~MB\\
			\hline I,II&19&20&0.27~MB&2.9~s&22~MB\\
			\hline I${}^+$&16&14&0.12~MB&1.9~s&16~MB\\
			\hline I${}^+$,II&13&14&0.12~MB&1.8~s&16~MB\\
			\hline
		\end{tabular}
	\end{table}
	
	\footnotetext{\label{FootnoteSpec}Total time and total memory
		means the amount of time and memory that is needed to solve the
		corresponding problem~\eqref{PLDE}; see~\eqref{Equ:Triangularized}
		and~\eqref{Equ:TriangularizedRef}; the time to set up this equation is almost constant and thus ignored. All the computations have been
		done with a standard notebook (11th Gen Intel® Core™ i7-1185G7 @ 3.00GHz × 8 with 16 GB memory) 
		using the computer algebra system Mathematica~13.0.}

	\begin{table}[h]
		\caption{\label{T:DoublePull}Double sum on the left hand side
			of~\eqref{Identity:ARK} (with preprocessing)}
		\begin{tabular}{|c|r@{$\times$}l|c|c|c|}\hline
			Improvements&equs&unknowns&size of system&total time$^{\ref{FootnoteSpec}}$& total memory$^{\ref{FootnoteSpec}}$\\
			\hline\hline None & 30&20&0.67~MB&5.7~s&28~MB\\
			\hline I& 21&17&0.21~MB&2.7~s&28~MB\\
			\hline II& 19&20&0.36~MB&3.3~s&29~MB\\
			\hline I,II&16&17&0.20~MB&2.4~s&29~MB\\
			\hline I${}^+$&16&14&0.12~MB&1.9~s&19~MB\\
			\hline I${}^+$,II&13&14&0.12~MB&1.9~s&19~MB\\
			\hline
		\end{tabular}
	\end{table}

	\begin{table}[h]
		\caption{\label{T:TripleAll}Triple sum~\eqref{Ahlgren:09AllInOne}
			(without preprocessing)}
		\begin{tabular}{|c|r@{$\times$}l|c|c|c|}\hline
			Improvements&equs&unknowns&size of system&total time$^{\ref{FootnoteSpec}}$& total memory$^{\ref{FootnoteSpec}}$\\
			\hline\hline None&81&47&12.0~MB&243 s& 100 MB\\
			\hline I&45&29&1.7~MB&28~s&43~MB\\
			\hline II&46&47&4.9~MB&67~s&45~MB\\
			\hline I,II&28&29&1.0~MB&13~s&34~MB\\
			\hline I${}^+$&24&20&0.45~MB&6.5~s&24~MB\\
			\hline I${}^+$,II&19&20&0.44~MB&6.5~s&24~MB\\
			\hline
		\end{tabular}
	\end{table}
	
	\begin{table}[h]
		\caption{\label{T:TriplePull}Triple sum in~\eqref{Ahlgren:09}
			(with preprocessing)}
		\begin{tabular}{|c|r@{$\times$}l|c|c|c|}\hline
			Improvements&equs&unknowns&size of system&total time$^{\ref{FootnoteSpec}}$& total memory$^{\ref{FootnoteSpec}}$\\
			\hline\hline None&52&31&3.2~MB&46~s&51~MB\\
			\hline I&37&25&1.1~MB&16~s&40~MB\\
			\hline II&30&31&1.4~MB&16~s&34~MB\\
			\hline I,II&24&25&1.0~MB&8.2~s&32~MB\\
			\hline I${}^+$&24&20&0.4~MB&5.6~s&27~MB\\
			\hline I${}^+$,II&19&20&0.4~MB&5.5~s&24~MB\\
			\hline
		\end{tabular}
	\end{table}
	\normalsize

	\subsection{Heuristic Check for the number of solutions}
	
	Usually, the field $\KK$ contains additional parameters like
	$\KK=\QQ(n)$, more generally say $\KK=\QQ(x_1,\dots,x_e)$. In this
	case, the bottleneck of the described algorithm is step {\bf(3)}.
	Suppose that we have computed a denominator bound
	$d(r)\in\KK[r]\setminus\{0\}$ and a degree bound $b\in\NN$ as described above.
	Then one can carry out the following speedups in step {\bf(3)}.
	
	\medskip
	
	Given $d(r)$
	and $b$, construct the linear system of equations with
	coefficients being polynomials in $\ZZ[x_1,\dots,x_e]$ as
	described in {\bf(3)}. Then, inspired
	by~\cite{LPS:02,RieseZimmermann:04}, we can check with inexpensive
	computations if a non-trivial solution for~\eqref{Equ:DenFreeLDE}
	and hence for~\eqref{PLDE} exists. More precisely, take a random
	prime number $p$ (sufficiently large) and random numbers $q_1,\dots,q_e$ from the
	finite field $\FF_p$ with $p$ elements. Afterwards, replace the
	parameters $x_i$ with $q_i$ in our linear equation system, and
	solve the system in the prime field $\FF_p$.

	\begin{Remark} Setting up the system with all the variables arising and afterwards performing
		the substitutions $x_i\mapsto q_i$ requires also a certain amount of computation time. Thus 
		we first carry out the substitution and afterwards derive the linear system with almost
		negligible effort.
	\end{Remark}

	Suppose that we find $s$ linearly independent solutions of the underlying system in the finite field $\FF_p$. Then the
	crucial observation is that there are at most $s$ solutions
	for~\eqref{Equ:DenFreeLDE} and thus for~\eqref{PLDE} in the original field $\KK(x)$; usually the determined $s$ agrees with the number of solutions for~\eqref{PLDE} up to some unlucky cases that we have never encountered so far.
	
	Hence with our check we obtain the following result:\\
	\noindent $\bullet$ If $s=0$, there are no non-trivial solutions
	for~\eqref{PLDE} and we can stop.\\
	$\bullet$ Otherwise, we conclude, or more precisely, suppose that
	there are exactly $s$ solutions for~\eqref{PLDE}; if there are less, we will discover this later. 
	With this information we proceed with our next improvement. 
	
\begin{Remark}
In general, one does not know (or is too lazy to predict) in advance  the order $\gamma$ for which a recurrence~\eqref{Equ:Recurrence} of the given double or multi-sum can be computed. One therefore starts with $\gamma=1$ (or even $\gamma=0$ in case a telescoping solution exist) and increments $\gamma$ step-wise until one finds the desired solution. In this regard, the heuristic check introduced above is extremely convenient to discover the non-existence of a solution without wasting too much calculation time. E.g., given the (hook-type) recurrences~\eqref{Rec:Triple:R} and~\eqref{Rec:Triple:RN}, it takes only 2.1 seconds to find out that our method fails to find a recurrence for the triple sum~\eqref{Ahlgren:09} by taking the instances $\gamma=0,1,2,3$.
\end{Remark}
	
	\subsection{Improvement I: Produce an optimal denominator and degree bound}
	
	Under the assumption that there exist exactly $s$ linearly independent solutions of~\eqref{PLDE}, we try to minimize the degree of the denominator bound
	$d(r)$ and to minimize the degree bound~$b$ as follows.
	Compute a complete factorization of
	$d(r)$ given in~\eqref{Equ:AbramovBound}, i.e.,
	\begin{equation}\label{Equ:DenBoundFactored}
		d(r)=d_1(r)^{m_1}\dots d_u(r)^{m_u}
	\end{equation}
	where the irreducible polynomials $d_i(r)$ occur with multiplicity
	$m_i>0$ in $d(r)$; as stated earlier, this can be done efficiently if the coefficients $a_0(r)$ and $a_{\delta}(r)$ in~\eqref{PLDE} are given already in factored form (which is usually the case). Then we test if also $d'(r):=d(r)/d_u(r)$ is a
	denominator bound by applying our {\bf Heuristic Check}
	for\footnote{Note that if $d'(r)$ is a denominator bound, also
		$b-1$ is a degree bound for the solutions
		of~\eqref{Equ:DenFreeLDE}. Namely, if we can reduce the degree of
		the ``denominator'' $d(r)$ by one we can also reduce the degree of
		the possible ``numerator'' by one.} $d'(r)$ and $b-1$; suppose
	that we have obtained $s'$ solutions. If $s$ is equal to $s'$,
	also $d'(r)$ is very likely a denominator bound for~\eqref{PLDE} --- except
	for some rare cases. In this case, one may go on with $d'(r)$ and $b-1$
	and cancel more and more factors $d_u(r)$ until the multiplicity
	$m_u$ of the factor $d_u(r)$ is zero or in our {\bf Heuristic
		Check} we get a different number of solutions than $s$.
	
	\begin{Remark}
		In \texttt{Sigma} this search is speeded up with a binary search tactic: First, we consider the multiplicity $\lambda=\lfloor m_u/2\rfloor$. If the number of solutions during our heuristic check remains $s$, we search recursively for the minimal multiplicity between
		$\lambda\in\{1,\dots,\lfloor m_u/2\rfloor\}$. Otherwise, we search within the range
		$\lambda\in\{\lfloor m_u/2\rfloor+1,\dots,m_u\}$.
	\end{Remark}

	In this way, we shall obtain an improved denominator bound, say
	$d_1(r)^{m_1}\dots d_{u-1}^{m_{u-1}}(r)\,d_u(r)^{\mu_u}$, where
	the multiplicity $\mu_u$ of the factor $d_u(r)$ is minimal.
	Analogously we proceed with the remaining irreducible factors.
	Finally, we arrive, up to unlucky cases, at a denominator bound,
	say $d'(r)$, whose degree is minimal.\\
	Next, we fix $d'(r)$ and reduce with the same tactic the degree bound $b$
	to $b'$ until our {\bf Heuristic Check} tells us that the degree
	bound $b'$ is minimal. Note that during this minimization process the number linearly
	independent solutions within the heuristic check always remains $s$.  
	
	\medskip
	
	To this end, we go on with step {\bf(3)} by using $d'(r)$ and
	$b'$. Suppose that we find $s'$ linearly independent solutions: If
	$s=s'$, we have found all solutions. Otherwise, only the situation $s'<s$ may arise, i.e., we might have lost some solutions. In
	this case, we repeat {\bf(3)} with the original denominator bound
	$d$ and degree bound $b$;  note that this (much more involved) situation never happened
	in our computations so far.

	\begin{Remark}For our applications in
		Sections~\ref{Sec:DoubleSums} and~\ref{Sec:MultipleSums} it suffices to find only one
		solution for~\eqref{PLDE}, i.e., we do not care if $s\neq s'$ as
		long as we get non-trivial solutions with $s'>0$.
	\end{Remark}
	
	\begin{Example}
		{\it Double sum on the left hand side of~\eqref{Identity:ARK}:}
		Given $d(r)$ from~\eqref{Equ:Kratt2Den:Pull:No} and $b=21$, we
		apply the {\bf Heuristic Check}. This test tells us that there are
		$s=1$ non-trivial solutions. Applying {\bf Improvement I} gives
		the {\it sharp} denominator bound
		$d'(r)=(n+1-r)^3(n+2-r)^3(n+3-r)^3$ and degree bound $b'=12$; for
		the
		properties of the linear system see the second row of Table~\ref{T:DoublePull}.\\
		With the same strategy we get the following results.\\
		{\it Double sum~\eqref{DoubleSumAhlgren7:InOne}:} We get the {\it
			sharp} bounds $d'(r)=(n-r)^3(n+1-r)^3(n+2-r)^3(n+3-r)^3$ and
		$b'=15$; for the properties of the linear system see
		the second row of Table~\ref{T:DoubleAll}.\\
		{\it Triple sum in~\eqref{Ahlgren:09}:} We get the {\it sharp}
		bounds $d'(r)=(n+1-r)^3(n+2-r)^3(n+3-r)^3(n+4-r)^3$ and $b'=19$;
		for the properties of the linear system see the second
		row of Table~\ref{T:TriplePull}.\\
		{\it Triple sum~\eqref{Ahlgren:09AllInOne}:} We get the {\it
			sharp} bounds
		$d'(r)=(n-1-r)^3(n-r)^3(n+1-r)^3(n+2-r)^3(n+3-r)^3(n+4-r)^3$,
		$b'=23$; for the properties of the linear system see the second
		row of Table~\ref{T:TripleAll}.
	\end{Example}
	
	\subsection{Improvement II: producing an optimal system}
	
	We observe that for all our linear systems with $u$ equations in $v$
	unknowns $u$ is much bigger as $v$ (without \textbf{Improvement I} it is almost twice as big). Under the assumption
	that there are $s>0$ linearly independent solutions it follows
	that $u-v-s$ equations can be removed. This observations leads us
	to remove step by step unnecessary
	equations. More precisely, we consider iteratively each equation
	and test if it can be removed without changing the solution set.
	If yes, we obtain a linear system with one equation less, and
	continue to check the remaining equations with this system.
	Otherwise, we go on without removing this equation.\\
	The crucial point is that this test can be carried out cheaply as
	follows. We test with our {\bf Heuristic Check} if the number of
	linearly independent solutions $s$ of the system, in which this
	equation is removed, equals to $s$. If yes, the set of solutions is the same
	--- up to some rare cases. Otherwise, we obtain more solutions
	($s'>s$), i.e., removing this equation is not possible. Following
	this strategy we obtain a linear system with $v-s$ equations in
	$v$ unknowns.\hfill\qed
	
	\medskip
	
	\noindent To this end, we may solve the reduced system of equations
	symbolically.

	\begin{Remark}
		The following two remarks are in place:\\
		$\bullet$ If one solves the system symbolically with
		Gauss-elimination, the unnecessary equations would have been
		eliminated implicitly -- but in a quite expensive manner.\\
		$\bullet$ Eliminating the unnecessary equations in different
		orders leads to different systems. In particular, there are
		tremendous differences in the time/memory behavior how these
		different systems can be solved symbolically. After testing
		various different strategies the following one turned out to be
		rather convincing: try to eliminate equations first which are
		given by the coefficients of lowest degree during the coefficient
		comparison in step {\bf(3)}.
	\end{Remark}
	
	\begin{Example}
		For the speedups using {\bf Improvement II}, we refer to the
		Tables~\ref{T:DoubleAll}--\ref{T:TriplePull}; more precisely, the
		entries in the third row (without {\bf Improvement I}) and the
		entries in the fourth row (together with {\bf Improvement I}).
	\end{Example}
	
	Summarizing, applying {\bf Improvements I} and {\bf II} in
	combination with {\bf Preprocessing} (pulling out factors of the
	multi-sum) can improve substantially our multi-sum method.
	
	\subsection{Improvement I${}^{+}$: predict contributions of the numerator solution}

	The proposed summation algorithms in Sections~\ref{Sec:DoubleSums} and~\ref{Sec:MultipleSums} start with the calculation of recurrences of univariate hypergeometric sums which can be also carried out, e.g., with the Paule-Schorn implementation~\cite{PauleSchorn:95} and its enhancements to deal with parameteried telescoping, as described in Section~\ref{Sec:ExistenceRec}. As it turns out, the specialized algorithms for univariate hypergeometric summation are still superior to our methods with all the above improvements. To understand this exceptional behavior, we note the following. Gosper's algorithm~\cite{Gosper:78,Paule:95,AequalB,CPS:08}, the backbone of the classical approach, relies on finding a rational solution $g(r)\in\KK(r)$ of the first-order linear recurrence
	\begin{equation}\label{Equ:GosperRatProblem}
		b(r)g(r+1)-g(r)=1
	\end{equation}
	where $t(r)$ is a hypergeometric term (usually built by a product of factorials, binomial coefficients, Pochhammer symbols) with $\frac{t(r+1)}{t(r)}=b(r)\in\KK(r)$. In order to find such a rational solution $g(r)$, the following steps are carried out~\cite{Gosper:78,AequalB,Paule:95,CPS:08}:\\
	\noindent{\bf(i)} One computes the Gosper representation of $b(r)$, i.e., non-zero polynomials $d(r),p(r),q(r)\in\KK[r]$ with
	$$b(r)=\frac{d(r+1)}{d(r)}\frac{p(r)}{q(r)}$$
	such that $\gcd(p(r),q(r+h))=1$ holds for all non-negative integers $h$.\\
	\noindent{\bf(ii)} Next, one decides constructively, if there exists a polynomial $\gamma(r)\in\KK[r]$ such that
	$$p(r)\,\gamma(r+1)-q(r-1)\,\gamma(r)=d(r)$$
	holds. Here one essentially proceeds as in our general procedure of step (2) given at the beginning of Section~\ref{Sec:SolveParaRec}.\\ 
	\noindent{\bf(ii)} If there is no $\gamma(r)\in\KK[r]$, then this implies that there is no $g(r)\in\KK(r)$ with~\eqref{Equ:GosperRatProblem}. Otherwise 
	one obtains the rational solution 
	\begin{equation}\label{Equ:GosperSol}
		g(r)=\frac{q(r)\gamma(r)}{d(r)}.
	\end{equation}
	In other words, $d(r)$ is a denominator bound of~\eqref{Equ:GosperRatProblem} and $g'(r)=q(r)\,\gamma(r)$ is the numerator contribution where $q(r)$ has been predicted by the Gosper ansatz. 
	This result can be further improved by refining the Gosper representation to computing the Gosper-Petkov{\v s}ek representation~\cite{Petkov:92,AequalB,CPS:08} in step (i) where in addition $\gcd(p(r),d(r))=\gcd(q(r),d(r+1))=1$ holds. As a consequence the predicted numerator contribution $q(r)$ in $g'(r)$ does not cancel with the denominator bound $d(r)$. Further, as elaborated, e.g., in~\cite{AequalB} this implies that among all possible choices of the Gosper representation, the degree $d(r)$ is minimal, i.e., we come close to a sharp denominator bound. This does not mean that there may still cancellations happen between $\gamma(r)$ and $d(r)$, but we have not found such an example so far. Comparing with our approach above, and knowing that we always obtain the optimal denominator bound $d(r)$, it is precisely the prediction of the numerator contribution $q(r)$ that makes Gosper's algorithm and all their variants, like Zeilberger's creative telescoping approach superior. A natural idea is to incorporate this extra feature to the general case to solve linear difference equations of the form~\eqref{PLDE}. This leads to

	\medskip
	
	\noindent\textbf{Improvement I${}^{+}$}: Given the recurrence~\eqref{PLDE} we set $b(r)=\frac{a_{\delta(r)}}{a_0(r)}\in\KK(r)$ and compute the non-zero polynomials $p(r),q(r),d(r)\in\KK[r]$ of  the generalized Gosper-Petkov{\v s}ek representation
	$$b(r)=\frac{d(r+\delta)}{d(r)}\frac{p(r)}{q(r)}$$
	where $\gcd(p(r),q(r+h\,\delta))=1$ holds for all non-negative integers $h$ and where, in addition, 
	$\gcd(p(r),d(r))=\gcd(q(r),d(r+\delta))=1$. This can be accomplished by the general algorithm presented in~\cite[Thm~2]{ABPS:21} for $\Pi\Sigma$-extensions; the calculation of the polynomial $a(r)$ can be skipped therein. In particular, we suppose that $q(r)$ is given in complete factorization, i.e.,
	\begin{equation}\label{Equ:PolyBoundFactored}
		q(r)=q_1(r)^{n_1}\dots q_u(r)^{n_u}
	\end{equation}
	where the irreducible polynomials $q_i(r)\in\KK[r]$ occur with multiplicity $n_i\in\NN$. 
	Now we proceed with step (1) but make the refined ansatz $g'(r)=q(r)\,\gamma(r)$ for some unknown polynomial $\gamma(r)$. Plugging $g'(r)$ into~\eqref{PLDE} yields
	\begin{align}\label{Equ:DenFreeLDEMod}
		\frac{a_{\delta}(r)q(r+\delta)}{d(r+\delta)}\,\gamma(r+\delta)+\dots+
		\frac{a_{1}(r)q(r+1)}{d(r+1)}\,\gamma(r+1)
		+\frac{a_0(r)q(r)}{d(r)}\,\gamma(r)=c_0\,f_0(r)+\dots+c_{\gamma}\,f_{\gamma}(r).
	\end{align}
	Next, we apply our heuristic check if there exist $s$ linearly independent solutions $\gamma(r)\in\KK[n]$. If not, we follow the strategy as in \textbf{Improvement~I} to find the maximal $n_i$ with $1\leq i\leq u$ such that all $s$ solutions can be recovered. Actually, we combine this technique with \textbf{Improvement I} and search simultaneously for the minimal $m_i\in\NN$ in~\eqref{Equ:DenBoundFactored}
	and the maximal $n_i$ in~\eqref{Equ:PolyBoundFactored} such that $g(r)=\frac{\gamma(r)q(r)}{d(r)}$ for some polynomial $\gamma(r)\in\KK[r]$ yields all solutions for \eqref{PLDE} or equivalently for~\eqref{Equ:DenFreeLDEMod}. In other words, in \textbf{Improvement I${}^{+}$} we search simultaneously for an optimal denominator bound~$d'(r)$ and try to predict extra factors of the numerator contribution, say $q'(r)$, together with the optimal degree bound $b'$ for the unknown contribution $\gamma(r)$ in $g=\frac{q'(r)\,\gamma(r)}{d'(r)}$.

	\begin{Remark}
		Restricting to the creative telescoping case of hypergeometric products, \textbf{Improvement I${}^{+}$} finds exactly the predicted factor $q(r)\in\KK[r]$ in~\eqref{Equ:GosperSol} of Gosper's method but guarantees also that $d(r)$ has minimal degree among all possible denominator bounds and that the degree bound of the unknown polynomial solution $\gamma(r)$ is minimal. In other words, it is the optimal ansatz that yields the same efficient behavior of all variants that utilize Gosper's algorithm; in some rare instances it may even outperform the Gosper-variants if the degree and denominator bounds of Gosper's method are not optimal. 
	\end{Remark}

	For general linear recurrences to determine the extra contribution $q(r)$ is a heuristic (in contrast to the very special first-order recurrence~\eqref{Equ:GosperRatProblem}) and one usually has to filter out wrong factors. Surprisingly enough, the found contributions are often non-trivial and contribute substantially to a speed up of our recurrence solver.
	
	\begin{Example}
		{\it Double sum on the left hand side of~\eqref{Identity:ARK}:}
		In order
		to compute~\eqref{Equ:Kratt2:pi} and~\eqref{Equ:Kratt2:g}, we compute not only the optimal denominator bound~\eqref{Equ:Kratt2Den:Pull:No} but also utilize the above tactic to find a non-trivial numerator contribution. More precisely, we obtain 
		$$q(r)=(2 n+1-r) (2 n+2-r) (r-1)^2 (r+1)^2$$
		and filter out wrong contributions yielding the correct factor 
		$$q'(r)=(2 n+1 -r) (r+1)^2$$
		of the solution~\eqref{Equ:Kratt2:g}. With this modified ansatz~\eqref{Equ:DenFreeLDEMod} the degree bound of $\gamma(r)$ is $b'=9$. This finally gives a linear system with 16 equations in 14 unknowns which require 0.12 MB of memory. Solving this system can be carried out in 1.9 seconds using 19MB of memory; compare row 5 in Table~\ref{T:DoublePull}.
		Applying in addition \textbf{Improvement II} enables one to eliminate three redundant constraints which leads basically to the same calculation time; compare row~6 in Table~\ref{T:DoublePull}.\\
		{\it Double sum~\eqref{DoubleSumAhlgren7:InOne}:} We get the numerator contribution $q'(r)=(2 n-r) (2 n+1-r) (r+1)^4$ and the degree bound $b'=9$ for the missing numerator part $\gamma(r)$; for the properties of the linear system see
		the 5th and 6th rows of Table~\ref{T:DoubleAll}.\\
		{\it Triple sum in~\eqref{Ahlgren:09}:} We find the numerator contribution 
		$q'(r)=(2 n+1-r) (r+1)^4$ together with the degree bound $b'=14$ for $\gamma(r)$;
		for the properties of the linear system see the 5th and 6th rows of  of Table~\ref{T:TriplePull}.\\
		{\it Triple sum~\eqref{Ahlgren:09AllInOne}:} We get the numerator factor $q'(r)=(2 n-r-1) (2 n-r) (2 n-r+1) (r+2)^6$ and the degree bound $b'=14$ for the unknown polynomial $\gamma(r)$; for the properties of the linear system see the 5th and 6th rows of Table~\ref{T:TripleAll}.
	\end{Example}
	
	The following general remarks for our proposed solving toolbox are in place.

	\begin{Remark}
		(a) If the number of solutions $s$ is larger than one, Improvements I and $I{}^+$ search for denominators $d\in\KK[r]$ and numerator contributions $q[r]$ that all solutions have in common. In particular, if $s=1$, this leads usually to much better bounds.\\
		(b) The above examples show that the derived linear system is almost optimal (see the 5th line of the tables) and \textbf{Improvement II} does not gain any further speedup (see the 6th line of the tables). Still this feature remains activated in order to deal with less optimal cases where the guess of the polynomial contributions in the numerator of the solution cannot be predicted sufficiently.
		We note further that for the special case $\KK=\QQ$ the linear system solver of Mathematica is so efficient that the gain to solve a system with the minimal number of rows is negligible. In this particular instance, \textbf{Improvement II} of \texttt{Sigma} is switched off. If more variables are contained in $\KK$, it is activated whenever the size of the input system is big enough to gain recognizable speed-ups.\\ 
		(c) With \texttt{Sigma} one can insert also manually\footnote{For the commands \texttt{GenerateRecurrence} and \texttt{SolveRecurrence} one can insert the additional option \texttt{UsePolynomialFactor$\to$p} to pass this factor $p(r)\in\KK[r]$ to the internal recurrence solver.} extra factors, say $p(r)\in\KK[r]$, which is merged with the automatically guessed factor $q'(r)$, i.e., $q'(r)$ is replaced by $\lcm(q'(r),p(r))\in\KK[r]$ and the above mechanism is activated.\\
		(d) In~\cite{vanHoeij:98} an improved version of Abramov's denominator bound algorithm has been introduced that finds a sharper denominator bound but can also provide some factors of the numerator. In all examples presented in this article van Hoeji's bound is exact, i.e., it is equivalent to our result after executing \textbf{Improvement I} or \textbf{Improvement I${}^+$}. Interestingly enough, our approach described as \textbf{Improvement I${}^+$} succeeds in finding substantially more numerator factors as the method proposed in~\cite{vanHoeij:98}. For instance, for the underlying recurrence~\eqref{PLDE} of the double sum 
		on the left hand side of~\eqref{Identity:ARK} the method from~\cite{vanHoeij:98} yields the numerator factor $(r+1)^2$ whereas we find the larger factor $(2 n+1 -r) (r+1)^2$. Similarly,
		for the double sum~\eqref{DoubleSumAhlgren7:InOne} the method of~\cite{vanHoeij:98} delivers no extra factor whereas our approach discovers the extra contribution $(2 n-r) (2 n+1-r) (r+1)^4$. Moreover, for the triple sum in~\eqref{Ahlgren:09} we find the numerator contribution 
		$(2 n+1-r) (r+1)^4$ whereas the method from~\cite{vanHoeij:98} delivers $(r+1)^4$. For all these cases the timings to solve the derived linear system (excluding the calculation time to execute the method in~\cite{vanHoeij:98}) is similar: it takes about 0.5 seconds more when one uses the bound from~\cite{vanHoeij:98}. 
		We conclude this observation by looking at the triple sum~\eqref{Ahlgren:09AllInOne}. Here we get the numerator factor $q'(r)=(2 n-r-1) (2 n-r) (2 n-r+1) (r+2)^6$ whereas the bound from~\cite{vanHoeij:98} is trivially $1$ and thus one obtains the same linear system given by \textbf{Improvement I}. In particular, solving this system needs $14$ seconds (instead of $7$ seconds using the factor $q'(r)$ of our approach).\\
		(e) Since the produced denominator bound in~\cite{vanHoeij:98} is rather good (e.g., it produces the optimal bound for all the recurrences under consideration), one may opt to use directly this bound and to optimize it further with our improvements. However the underlying algorithm is much more time consuming than applying the formula~\eqref{Equ:DenBoundFactored} with all our improvements. E.g., computing~\eqref{Equ:AbramovBound}, producing the factored form~\eqref{Equ:DenBoundFactored} and minimizing the multiplicities $m_i$ takes less than 0.5 seconds in all our examples. But carrying out the method in~\cite{vanHoeij:98} takes with our (maybe not optimal) implementation longer than solving the whole system. Thus we permanently switched off the approach given in~\cite{vanHoeij:98} in the summation package \texttt{Sigma} and take as starting point for our simplification the formula~\eqref{Equ:DenBoundFactored}.\\
		(f) All the improvements carry over straightforwardly to the $q$-case, i.e., where the coefficients $a_i(r)$ in~\eqref{PLDE} are from $\KK[q^r]$ over the rational function field $\KK=\KK'(q)$ and one looks for all solutions in the rational function field $\KK(q^r)$.\\
		(g) The described recurrence solver with all its improvements is not only the backbone for the multi-sum approach but is also an important key ingredient of the \texttt{Sigma} function \texttt{SolveRecurrence} to find efficiently 
		all d'Alembertian solutions~\cite{Abramov:94,AequalB,ABPS:21} over $\Pi\Sigma$-fields.
	\end{Remark}

\section{Conclusion}\label{Sec:Conclusion}
We presented a fast method to compute linear recurrences for hypergeometric double sums that is also suitable for multiple sums. To guarantee the success of this method, the algorithmic theory of contiguous relations has been exploited. In addition new ideas have been presented to find rational solutions of parameterized recurrences efficiently. All the algorithmic ideas of our summation method and also the improvements of the recurrence solving extend in a straightforward fashion to the $q$-hypergeometric case and are available within \texttt{Sigma}.


\begin{thebibliography}{BBF{\etalchar{+}}14}
	
	\bibitem[ABPS21]{ABPS:21}
	S.A. Abramov, M. Bronstein, M. Petkov{\v s}ek, and C. Schneider.
	\newblock {On rational and hypergeometric solutions of linear ordinary
		difference equations in $\Pi\Sigma^*$-field extensions}.
	\newblock {\em J. Symb. Comput.}, 107:23--66, 2021.
	\newblock arXiv:2005.04944 [cs.SC].
	
	\bibitem[Abr89a]{Abramov:89b}
	S.A. Abramov.
	\newblock Problems in computer algebra that are connected with a search for
	polynomial solutions of linear differential and difference equations.
	\newblock {\em Moscow Univ. Comput. Math. Cybernet.}, 3:63--68, 1989.
	
	\bibitem[Abr89b]{Abramov:89a}
	S.A. Abramov.
	\newblock Rational solutions of linear differential and difference equations
	with polynomial coefficients.
	\newblock {\em U.S.S.R. Comput. Math. Math. Phys.}, 29(6):7--12, 1989.
	
	\bibitem[Abr95]{Abramov:95a}
	S.A. Abramov.
	\newblock Rational solutions of linear difference and $q$-difference equations
	with polynomial coefficients.
	\newblock In T.~Levelt, editor, {\em Proc. ISSAC'95}, pages 285--289. ACM
	Press, New York, 1995.
	
	\bibitem[ABRS12]{ABRS:12}
	J.~Ablinger, J.~Bl\"umlein, M.~Round, and C.~Schneider.
	\newblock Advanced computer algebra algorithms for the expansion of Feynman
	integrals.
	\newblock In J.~Bluemlein, S.~Moch, and T.~Riemann, editors, {\em {Loops and
			Legs in Quantum Field Theory 2012}}, volume~50 of {\em PoS(LL2012)}, pages
	1--14, 2012.
	
	\bibitem[AP93]{AndrewsPaule:93}
	G.E. Andrews and P.~Paule.
	\newblock Some questions concerning computer-generated proofs of a binomial
	double-sum identity.
	\newblock {\em J.~Symbolic Comput.}, 16:147--153, 1993.
	
	\bibitem[AP94]{Abramov:94}
	S.A. Abramov and M.~Petkov{\v s}ek.
	\newblock D'{A}lembertian solutions of linear differential and difference
	equations.
	\newblock In J.~von~zur Gathen, editor, {\em Proc. ISSAC'94}, pages 169--174.
	ACM Press, Baltimore, 1994.
	
	\bibitem[APS05]{AndrewsPauleSchneider:04b}
	G.E. Andrews, P.~Paule, and C.~Schneider.
	\newblock Plane partitions {V}{I}: {S}tembridge's {T}{S}{P}{P} {T}heorem.
	\newblock {\em Advances in Applied Math. Special Issue Dedicated to Dr. David
		P. Robbins. Edited by D. Bressoud}, 34(4):709--739, 2005.
	\newblock Preliminary version online.
	
	\bibitem[BBF{\etalchar{+}}14]{BBFPRRAHSWM:14}
	A.~Behring, J.~Bl\"umlein, A.~De Freitas, T.~Pfoh, C.~Raab, M.~Round,
	J.~Ablinger, A.~Hasselhuhn, C.~Schneider, F.~Wissbrock, and A.~von
	Manteuffel.
	\newblock New results on the 3-loop heavy flavor corrections in deep-inelastic
	scattering.
	\newblock In {\em {Proc. of RADCOR}}, PoS(RADCOR 2013)058, pages 1--21, 2014.
	
	\bibitem[BRS18]{BRS:18}
	J.~Bl\"umlein, M.~Round, and C.~Schneider.
	\newblock Refined holonomic summation algorithms in particle physics.
	\newblock In E.~Zima and C.~Schneider, editors, {\em { Advances in Computer
			Algebra. WWCA 2016.}}, volume 226 of {\em Springer Proceedings in Mathematics
		\& Statistics}, pages 51--91. Springer, 2018.
	\newblock arXiv:1706.03677 [cs.SC].
	
	\bibitem[Chy00]{Chyzak:00}
	F.~Chyzak.
	\newblock An extension of {Z}eilberger's fast algorithm to general holonomic
	functions.
	\newblock {\em Discrete Math.}, 217:115--134, 2000.
	
	\bibitem[CPS08]{CPS:08}
	W.~Y. Chen, P. Paule, and H.L. Saad.
	\newblock Converging to {G}osper's algorithm.
	\newblock {\em Adv. in Appl. Math.}, 41(3):351--364, 2008.
	
	\bibitem[Gau13]{Gauss}
	C.F. Gau{\ss}.
	\newblock Disquisitiones generales circa seriem infinitam $1+\frac{\alpha
		\beta}{1 \cdot \gamma} x +\dots$, pars prior.
	\newblock {\em Commentationes societatis regiae scientarum Gottingensis
		recentiores 2 (classis
		mathematicae)},\url{https://gdz.sub.uni-goettingen.de/id/PPN235999628}, 1813.
	
	\bibitem[Gos78]{Gosper:78}
	R.W. Gosper.
	\newblock Decision procedures for indefinite hypergeometric summation.
	\newblock {\em Proc. Nat. Acad. Sci. U.S.A.}, 75:40--42, 1978.
	
	\bibitem[Kar81]{Karr:81}
	M.~Karr.
	\newblock Summation in finite terms.
	\newblock {\em J.~ACM}, 28:305--350, 1981.
	
	\bibitem[KR04]{Krattenthaler:04}
	C.~Krattenthaler and T.~Rivoal.
	\newblock Hyperg\'{e}om\'{e}trie et fonction z\^{e}ta de {R}iemann.
	\newblock {\em Preprint}, 2004.


\bibitem[MS18]{LPS:02}
	R. Lyons, P. Paule, and A. Riese.
	\newblock A computer proof of a series evaluation in terms of harmonic numbers.
	\newblock Appl. Algebra Engrg. Comm. Comput., 13:327--333, 2002.
	
	\bibitem[MS18]{MS:18}
	J.~Middeke and C.~Schneider.
	\newblock {Denominator bounds for systems of recurrence equations using
		$\Pi\Sigma$-extensions}.
	\newblock In C.~Schneider and E.~Zima, editors, {\em { Advances in Computer
			Algebra. WWCA 2016.}}, volume 226 of {\em Springer Proceedings in Mathematics
		\& Statistics}, pages 149--173. Springer, 2018.
	\newblock arXiv:1705.00280 [cs.SC].
	
	\bibitem[Pau95]{Paule:95}
	P.~Paule.
	\newblock Greatest factorial factorization and symbolic summation.
	\newblock {\em J.~Symbolic Comput.}, 20(3):235--268, 1995.
	
	\bibitem[Pau21]{Paule:21}
	P.~Paule.
	\newblock {Contiguous relations and creative telescoping}.
	\newblock In J.~Bl\"umlein and C.~Schneider, editors, {\em {Anti-Differentiation
			and the Calculation of Feynman Amplitudes}}, Texts and Monographs in Symbolic
	Computation, pages 335--394. Springer, 2021.
	
	\bibitem[Pet92]{Petkov:92}
	M.~Petkov{\v s}ek.
	\newblock Hypergeometric solutions of linear recurrences with polynomial
	coefficients.
	\newblock {\em J.~Symbolic Comput.}, 14(2-3):243--264, 1992.

	
	\bibitem[PS95]{PauleSchorn:95}
	P.~Paule and M.~Schorn.
	\newblock A {M}athematica version of {Z}eilberger's algorithm for proving
	binomial coefficient identities.
	\newblock {\em J.~Symbolic Comput.}, 20(5-6):673--698, 1995.
	
	\bibitem[PS03]{PauleSchneider:03}
	P.~Paule and C.~Schneider.
	\newblock Computer proofs of a new family of harmonic number identities.
	\newblock {\em Adv. in Appl. Math.}, 31(2):359--378, 2003.
	
	\bibitem[PW00]{PetkovWeix}
	M.~Petkov{\v s}ek and C.~Weixlbaumer.
	\newblock A comparison of degree polynomials.
	\newblock \url{http://www.fmf.uni-lj.si/~petkovsek/}, 2000.
	\newblock Note.
	
	\bibitem[PWZ96]{AequalB}
	M.~Petkov{\v s}ek, H.S. Wilf, and D.~Zeilberger.
	\newblock {\em $A=B$}.
	\newblock A.K. Peters, Wellesley, MA, 1996.
	
	\bibitem[RZ04]{RieseZimmermann:04}
	A.~Riese and B.~Zimmermann.
	\newblock Randomization speeds up hypergeometric summation.
	\newblock {\em Preprint}, 2004.
	
	\bibitem[SA95]{Abramov:95b}
	M.~Petkov{\v s}ek S.A.~Abramov, M.~Bronstein.
	\newblock On polynomial solutions of linear operator equations.
	\newblock In T.~Levelt, editor, {\em Proc. ISSAC'95}, pages 290--296. ACM
	Press, New York, 1995.
	
	\bibitem[Sch01]{Schneider:01a}
	C.~Schneider.
	\newblock {\em Symbolic Summation in Difference Fields}.
	\newblock PhD thesis, RISC-Linz, J.~Kepler University, Linz, May 2001.
	
	\bibitem[Sch05]{Schneider:04c}
	C.~Schneider.
	\newblock A new {S}igma approach to multi-summation.
	\newblock {\em Advances in Applied Math. Special Issue Dedicated to Dr. David
		P. Robbins. Edited by D. Bressoud}, 34(4):740--767, 2005.
	\newblock Preliminary version online.
	
	\bibitem[Sch07]{Schneider:07}
	C.~Schneider.
	\newblock {Symbolic Summation Assists Combinatorics}.
	\newblock {\em Sem.~Lothar. Combin.}, 56:1--36, 2007.
	\newblock Article B56b.
	
	\bibitem[Sch16]{Schneider:16}
	C.~Schneider.
	\newblock A difference ring theory for symbolic summation.
	\newblock {\em J. Symb. Comput.}, 72:82--127, 2016.
	
	\bibitem[Str94]{Strehl:94}
	V.~Strehl.
	\newblock Binomial identities --- combinatorial and algorithmical aspects.
	\newblock {\em Discrete Math.}, 136:309--346, 1994.
	
	\bibitem[SZ21]{SZ:21}
	C.~Schneider and W.~Zudilin.
	\newblock {A case study for $\zeta(4)$}.
	\newblock In Alin Bostan and Kilian Raschel, editors, {\em {Transcendence in
			Algebra, Combinatorics, Geometry and Number Theory. TRANS 2019}}, volume 373
	of {\em Proceedings in Mathematics \& Statistics}, pages 421--435. Springer,
	2021.
	\newblock arXiv:2004.08158 [math.NT].
	
	\bibitem[vH98]{vanHoeij:98}
	M.~van Hoeij.
	\newblock Rational solutions of linear difference equations.
	\newblock In O.~Gloor, editor, {\em Proc. ISSAC'98}, pages 120--123. ACM Press,
	1998.
	
	\bibitem[Weg97]{Wegschaider:97}
	K.~Wegschaider.
	\newblock Computer generated proofs of binomial multi-sum identities.
	\newblock Di\-plo\-ma thesis, RISC Linz, Johannes Kepler University, May 1997.
	
	\bibitem[Zei90]{Zeilberger:90a}
	D.~Zeilberger.
	\newblock A holonomic systems approach to special functions identities.
	\newblock {\em J. Comput. Appl. Math.}, 32:321--368, 1990.
	
\end{thebibliography}

\newcommand{\etalchar}[1]{$^{#1}$}

\end{document}